\documentclass[fleqn,usenatbib]{mnras}

\usepackage{newtxtext,newtxmath}
\usepackage{multicol}
\usepackage{multirow}
\usepackage{booktabs}
\usepackage{multirow}
\usepackage{siunitx}
\usepackage[T1]{fontenc}

\DeclareRobustCommand{\VAN}[3]{#2}
\let\VANthebibliography\thebibliography
\def\thebibliography{\DeclareRobustCommand{\VAN}[3]{##3}\VANthebibliography}

\usepackage{graphicx}	% Including figure files
\usepackage{amsmath}	% Advanced maths commands
\title[TOI-1338 and TIC-172900988 orbits]{Orbits of the TOI-1338 and TIC-172900988 systems}

\author[Dionysios Gakis \& Konstantinos N. Gourgouliatos]{
Dionysios Gakis$^{1}$\thanks{E-mail: dgakis@upnet.gr}
\& Konstantinos N. Gourgouliatos$^{1}$\thanks{E-mail: kngourg@upatras.gr}
\\
$^{1}$Department of Physics, University of Patras, Patras, Rio, 26504, Greece
}

\date{Accepted 2022 December 19. Received 2022 December 19; in original form 2022 July 12}

\pubyear{2022}

\begin{document}
\raggedbottom
\label{firstpage}
\pagerange{\pageref{firstpage}--\pageref{lastpage}}
\maketitle

\begin{abstract}
Recent observations by TESS revealed the existence of circumbinary planets in the systems of TOI-1338 and TIC-172900988. The purpose of this work is to model the planetary orbits in these two systems and study them under the perspective of previous theoretical models. Each planet's distance from the barycenter through time is simulated using n-body integrations and is compared with outcomes from a semi-analytic, a geometric and a Keplerian-based approach. Furthermore, we infer the most prominent frequencies of both planets' orbits induced by the central binaries. We confirm that both systems appear to be stable. Lastly, we examine the implications of an additional candidate planet in TOI-1338 system finding that an extra, 48 $M_{\earth}$ planet that has been hinted from observations could be located at $0.8$ AU without generating any radical changes to the orbits of the other members of the system.
\end{abstract}

\begin{keywords}
celestial mechanics -- stars: planetary systems
\end{keywords}

%%%%%%%%%%%%%%%%%%%%%%%%%%%%%%%%%%%%%%%%%%%%%%%%%%

%%%%%%%%%%%%%%%%% BODY OF PAPER %%%%%%%%%%%%%%%%%%

\section{Introduction}

Binary stelllar systems are thought to be abundant in the universe. Hence, theoretical calculations arising by analyzing statistical data suggest that at least roughly half of the the Milky Way stars should be multiple \citep{Duquennoy:1991,Kouwenhoven:2007,Raghavan:2010,Duchene:2013,Tokovinin:2014}. Therefore, circumbinary exoplanets are rather common, even if conservatively assuming that planetary creation around binary stars does not have a high probability \cite[e.g.][]{Moriwaki:2004,Scholl:2007,Pierens:2007,Bromley:2015}. We find double systems in the Solar System too, the most notable example being the Pluto-Charon binary dwarf planet \citep{Buie:2006} and its four surrounding small moons \citep{Weaver:2006, Showalter:2011, Showalter:2012}. 

Yet, despite their alleged plethora \citep{Armstrong:2014,Martin:2014} and after being long foreseen \cite[e.g.][]{Schneider:1994}, exoplanets around double stars were only identified for the fist time during the previous decade. The first discovery of a circumbinary planet was in 2011, when Kepler-16b was detected by \cite{Doyle:2011}. Kepler-16b has a mass similar to Saturn's and is orbiting around two low-mass main sequence stars. \cite{Welsh:2012} discovered two additional circumbinary exoplanets, named Kepler-34 b and Kepler-35 b, also found from observations by the Kepler mission \citep{Borucki:2011}. Several other planets of this category were also discovered by Kepler \citep{Orosz:2012a,Orosz:2012b,Schwamb:2013,Kostov:2013,Kostov:2014,Welsh:2015,Kostov:2016,Orosz:2019,Socia:2020}. 

The next generation of telescopes aiming to find and characterize exoplanets came in 2018, with the NASA's Transiting Exoplanet Survey Satellite (TESS) space telescope \citep{Howard:2015}. TESS photometric data resulted in the discovery of another two circumbinary planets, TOI-1338 \citep{Kostov:2020} and TIC-172900988 \citep{Kostov:2021} systems, which are the the topic of this study. The number of discovered exoplanets around binary systems is constantly increasing, currently being around 20. Up until now, all these planets are almost co-planar with the central stellar orbits, which is perhaps indicative of the dynamical situation of such systems, or a bias effect of the observations \citep{Li:2016}. Moreover, if we compare them to the only nearby system that we have been able to study in detail, the Pluto-Charo binary dwarf planet, they all differ in the fact that the central binary orbits have a considerable eccentricity, which is not the case for the Pluto-Charon system, whose orbit is almost circular.

TOI-1338 circumbinary planet was identified by \cite{Kostov:2020} orbiting the eclipsing binary star EBLM J0608-59. Some months later, the discovery of the second circumbinary exoplanet by TESS observations, around TIC-172900988 double system, was announced \citep{Kostov:2021}. The orbits of the circumbinary planets were also first-established in the above studies. An analysis of the dynamics within the TOI-1338 and TIC-172900988 systems has been given by \cite{Georgakarakos:2022}. In that 
paper, a fictitious Earth-like planet was placed in the habitable zone around both of the binary systems and the long-term evolution of the two systems, now consisting of two circumbinary exoplanets, was inspected. The author found that such an additional planet could not unsettle the dynamical situation.

In general, the study of the dynamical architecture of circumbinary (or multiple) systems seems to be of particular interest. For example, \cite{Woo:2018}, \cite{Bromley:2020a} studied the secular evolution of the Pluto and Charon and compared their final outcomes with the present-day conditions, in order to determine whether their moon system could be formed after a large collision or through in-situ accretion. Similarly, there has been a number of studies, studying typical circumbinary disks, aiming to assess the role of alignment \cite[e.g.][]{Childs:2022} or resonances \cite[e.g.][]{Sutherland:2019} in the formation of circumbinary planets. Another useful application of the orbital analysis of orbits around binaries, is the search for other habitable zones \cite[e.g.][]{Muller:2014}.

In a previous work, \citep{Gakis:2022}, we studied the small moon motions around the binary dwarf planet system of Pluto and Charon. Our analysis indicated that systems like the Pluto-Charon one may well be approximated by both semi-analytic and numerical models. The results obtained were compatible between the two methods, giving only a maximum deviation of 0.2925\%, which may be attributed to the higher order terms neglected by the linearized model, and more importantly to the mutual gravitational interactions caused by each other moons. These reciprocal effects were found to be rather significant in the low-frequency region. 

Here, we apply similar techniques in the recently discovered circumbinary exoplanets by TESS, TOI-1338 and TIC-172900988. Our aim is to give a
thorough perspective on the orbital behavior of the above systems. We implement a series of proposed techniques to keep track on the circumbinary properties through time, and examine the possibility of the existence of additional objects, not yet revealed by observations. At the moment this paper is being written, the two systems are known to host a single planet, but there are indications that other objects might be present, such as second planet in TOI-1338 \citep{Standing:2021}. The stability of both systems is discussed and comparisons between them are being made.

The structure of the paper is the following: after briefly describing the dynamical properties of the in-question circumbinary systems (Section \ref{sec:sec2}), we outline the methods we apply for estimating circumbinary orbital characteristics (Section \ref{sec:sec3}). Then, we present our results (Section \ref{sec:sec4}) and discuss their implications in Section \ref{sec:sec5}. We conclude and summarize our work in Section \ref{sec:sec6}. 

\section{Dynamical properties of circumbinary exoplanets} \label{sec:sec2}

The typical dynamical properties of TOI-1338 and TIC-172900988 have been determined in the discovery papers, \cite{Kostov:2020} and \cite{Kostov:2021}, respectively. We note though that the relevant discussion therein is limited to identifying the standard Keplerian orbital frequency, along with the nodal and apsidal frequency in their frequency spectrum, without any analysis of the large number of the frequencies at which the circumbinary planets oscillate, their various oscillatory amplitudes or an attempt to model their orbits under a mathematical formulation. This is the objective of our study. In this section, we outline the dynamical results by \cite{Kostov:2020} and \cite{Kostov:2021} for TOI-1338 and TIC-172900988, respectively, which act as the initial parameters for our analysis.

\subsection{TOI-1338}
\label{sec:TOI} % used for referring to this section from elsewhere

At the moment, the TOI-1338 system is considered as a hierarchical 3-body system, consisting of a giant planet following a nearly circular (e $\approx$ 0.09) P-type orbit around the binary star in 95.2 days. The binary orbit has a small eccentricity (e $\approx$ 0.16) and a period of 14.6 days, while the %total 
system has an inclination of around 1$^{\circ}$. The host stars have masses 1.1 $M_{\sun}$ and 3.3 $M_{\sun}$ and radii 1.3 $R_{\sun}$ and 0.3 $R_{\sun}$, respectively \citep{Triaud:2017}. In terms of the planetary properties, the planet, at a mass of 33.0 $\pm$ 20.0 $M_{\earth}$, radius of 6.85 $\pm$ 0.19 $R_{\earth}$ and a bulk density of 0.56 $\pm$ 0.34 g cm$^{-3}$, is analogous to Saturn.

The above values, among others, were estimated by \cite{Kostov:2020}. The photometric data by TESS were combined with ground-based precise radial velocity measurements from CORALIE and HARPS, and a complete photometric-dynamical model was studied to determine the parameters of the system. Hence, using n-body integrations (mainly Newtonian, but also including general relativity and tidal evolution effects when necessary) and observations of eclipses and transits, a set of 25 parameters was solved for. The main orbital parameters of the TOI-1338 system, as calculated by \cite{Kostov:2020}, are presented in the upper part of Table \ref{tab:1}.

The orbit of the planet lies in between the 6:1 and 7:1 mean motions resonances (MMRs) with the binary \citep{Kostov:2020}. Furthermore, the authors concluded that the semi-major axis of the planet is approximately 30\% larger the critical distance for instability of the analysis by \cite{Holman:1999} and its extension, \citep{Quarles:2018}. The same outcome was shown by \cite{Georgakarakos:2022}, adopting the model by \cite{Georgakarakos:2015} for circular orbits. Consequently, TOI-1338 is validated as a stable system. 

\subsection{TIC-172900988}
\label{sec:TIC}

The study of the characteristics of TIC-172900988 was given by \cite{Kostov:2021}. In this case, the eclipsing binary stars revolve around each other in $\approx$ 19.7 days in quite
eccentric orbits (e $\approx$ 0.45). The two stars have similar sizes; the primary has a mass of 1.24 $M_{\sun}$ and a radius of 1.38 $R_{\sun}$, whereas the corresponding values for the secondary star are 1.20 $M_{\sun}$ and 1.30 $R_{\sun}$. All of these values, together with the orbital parameters of the hosted planet, as provided by \cite{Kostov:2021}, are summarized in the upper part of Table \ref{tab:2}.

Nevertheless, the orbital characteristics of the planet could not be uniquely determined. The analysis performed in the above work resulted in six equal-probability families of solutions for the mass and orbit of the planet. Thus, the orbital period of the circumbinary planet ranges between 190 and 205 days, and its mass lies in the interval of 824-981 $M_{\earth}$. Future observations may clarify which one is the correct solution. Hereafter, we adopt the Family 5 solution for our calculations, as considered the most accurate one \citep{Kostov:2021}. So, the Family 5 solution is given for the planetary orbital elements in Table \ref{tab:2}.

The mean distance of the planet from the system's barycenter is 34\% larger than the critical distance for instabilities \citep{Holman:1999,Quarles:2018}. As a result, the system is classified as stable. This outcome is validated by long-term 3-body integrations by \citep{Kostov:2021}, for all of the six families, which showed no significant rise of the eccentricity through time.

\section{Methods for circumbinary orbit characterization }
\label{sec:sec3}

The most prominent work of categorizing ways to model and quantify circumbinary orbits is the one by \cite{Bromley:2020}. In this paper, the authors develop several methods to distinguish the random motion of a circumbinary orbit from the forced oscillations by the binary system. They test their ideas by applying them to the Kepler-16, Kepler-47, and Pluto–Charon systems. In this section, we describe the theoretical methods we use to quantify the circumbinary orbits in TOI-1338 and TIC-172900988. By investigating several models for circumbinary orbits, we can compare the accuracy of each one of them and adopt the most suitable ones in later studies.

\subsection{Linearized model} \label{subsec:linearized}

In this section, a short review of the semi-analytic model used in this work is given, which is the theory developed by \cite{Leung:2013}. This model is in fact a generalization of the \cite{Lee:2006} model, now including eccentric binaries as well, while the latter accounted for central orbits of zero (or negligible) eccentricity. The orbital eccentricity of the TOI-1338 host binary has been calculated to $e=0.16$, while the respective value for TIC-172900988 is a lot larger, $e=0.45$. It comes as no surprise then that the \cite{Leung:2013} model is more suitable for our analysis. Besides, the authors of this paper collated their theoretical solutions with outcomes by numerical integrations of the Kepler-16 b, Kepler-34 b, and Kepler-35 b circumbinary exoplanets, finding sufficient convergence.

At first, we define a set of cylindrical coordinates, so that the point $O\,(0,0,0)$ represents the barycenter of the system. The position vectors of the two stars comprising the binary system, A and B, are $\mathbf{r_A}=(a_A, \phi _B+\pi, 0)$ and $\mathbf{r_B}=(a_B, \phi _B, 0)$, where $a_A=a_{bin}\,m_{B}/m_{bin}$, $a_B=a_{bin}\,m_{A}/m_{bin}$, $a_{bin}=a_A + a_B$, $m_{bin}=m_A + m_B$ and $\phi _B(t)=n_{bin}t+\phi '$ ($\phi '$ is a constant). A and B revolve around their common center of mass at a circular frequency of $n_{bin}=[G(m_{A}+m_{B})/a_{bin}^3]^{1/2}$ . At a point P$(R, \phi, z)$, the potential is:
\begin{equation}\label{1} 
\Phi(R, \phi, z) = -\frac{Gm_A}{|\mathbf{r}-\mathbf{r_A}|}-\frac{Gm_B}{|\mathbf{r}-\mathbf{r_B}|}\,.
\end{equation}
Expanding the distances of the AP and BP into a cosine series, as in \cite{Murray:1999}, the above equation may be written as: 
\begin{eqnarray}
\Phi(R, \phi, z=0) = \sum_{k=0}^{\infty} \Phi_{k0} \cos k(\phi-M_B-\varpi_B)
\label{Phi} \\ 
+ e_{AB} \sum_{k=0}^{\infty} \left( k\Phi_{k0} - {1 \over 2} \Phi_{k1}\right) \cos (k(\phi-\varpi_B)-(k+1) M_B) \nonumber \\ 
+ e_{AB} \sum_{k=0}^{\infty} \left(-k\Phi_{k0}  - {1 \over 2} \Phi_{k1}\right) \cos
(k(\phi-\varpi_B)-(k-1) M_B), \nonumber
\end{eqnarray}
where $\Phi_{k0}$ and $\Phi_{k1}$ are functions of the binary masses, their separation and the orbital radius of the planet, $e_{AB}$is the eccentricity of the binary system and $M_B = n_{bin} t + \varphi_{bin}$ is the mean anomaly of B relative to A ($\varphi_{bin}$ is a constant). 

The three equations of motion are:
\begin{equation}\label{3} 
\frac{d^2R}{dt^2}-R\left(\frac{d\phi}{dt}\right)^{2}=-\frac{\partial \Phi}{\partial R}\,,
\end{equation}
\begin{equation}\label{4} 
R\frac{d^2\phi}{dt^2}+2 \frac{R}{dt} \frac{\phi}{dt}=-\frac{1}{R}\frac{\partial \Phi}{\partial \phi}\,,
\end{equation}
\begin{equation}\label{5} 
\frac{d^2z}{dt^2}=-\frac{\partial \Phi}{\partial z}\,,
\end{equation}
The above system has been worked out by \cite{Leung:2013} and the result yields the following solution:
\begin{eqnarray}\label{6} 
R(t) &=&
R_P \Bigg\{1 - e_{\rm free} \cos(v_e t + \kappa) - C_0 \cos(n_{bin} t + \nu) \nonumber\\
& &- \sum_{k=1}^{\infty} \Big[C_k^0 \cos k(n_{S}t  -\varpi_B - n_{bin} t -\nu) \nonumber\\
& & + C_k^+ \cos (k(n_{S}t-\varpi_B)-(k+1)(n_{bin} t +\nu)) \nonumber\\
& &+ C_k^- \cos (k(n_{S}t-\varpi_B)-(k-1)(n_{bin} t +\nu))
\Big]
\Bigg\} \,, 
\end{eqnarray}
\begin{eqnarray}
\phi(t) &=&
 n_{syn} t 
+ {2 n_{syn} \over v_e} e_{\rm free} \sin(v_e t + \kappa)
+ {n_{syn} \over n_{bin}} D_0 \sin (n_{bin} t +\nu)
\nonumber\\
& & + \sum_{k=1}^{\infty} \Big[
{n_{syn} \over k (n_{syn} - n_{bin})} D_k^0 \sin k(n_{syn}t-\varpi_B - n_{bin} t -\nu))
\nonumber\\
& & + {n_{syn} \over k n_{syn} - (k+1) n_{bin}} D_k^+ \sin (k(n_{syn}t-\varpi_B)\nonumber\\
& &-(k+1)(n_{bin} t +\nu))
\nonumber\\
& & + {n_{syn} \over k n_{syn} - (k-1) n_{bin}} D_k^- \sin (k(n_{syn}t-\varpi_B)\nonumber\\
& &-(k-1)(n_{bin} t +\nu))
\Big]\,,
\label{phit}
\end{eqnarray}
\begin{equation}\label{8} z(t)=iR_P\cos \left(v_it+\lambda\right)\,, 
\end{equation}
where the coefficients $C_0$, $C_k^0$, $C_k^\pm$ and $D_k$ represent the mode amplitudes, and $\kappa$, $\nu$ and $\lambda$ are constants. The detailed proof may be found in \cite{Leung:2013}. The calculated values of the coefficients  $C_0$, $C_k^0$, $C_k^\pm$ (up until $k = 3$) for TOI-1338 and TIC-172900988 exoplanets are given in Table \ref{tab:3}. Free eccentricity $e_{free}$ is an independent parameter in the solution and represents the amplitude of the epicyclic motion, around a mean distance, as quantified by $R_P$. The longitude of periapsis of B relative to A is $\varpi_B = \Omega_B + \omega_B$. $n_{syn}$ stands for the synodic frequency, i.e. $n_{syn}=n_{bin}-n_S$ and $i$ is the inclination of the planet orbit with respect to the binary's orbital plane. The planet's mean motion $n_{S}$, the epicyclic frequency $v_e$ and the vertical frequency $v_i$ are defined as follows:
\begin{equation} n_{S}^2=\frac{1}{R_S}\frac{d\Phi _{000}}{dR}\Bigg|_{R_S} \,, \end{equation}
\begin{equation} 
v_{e}^2=R_S\frac{dn _{S}^2}{dR}\Bigg|_{R_S}+4n _{S}^2\,, \end{equation}
\begin{equation} 
v_{i}^2=\frac{1}{z}\frac{d\Phi }{dz}\Bigg|_{z=0,\, R_S}\,. \end{equation}
The final expressions for the orbital coordinates are very similar in the two semi-analytic theories of \cite{Lee:2006} and \cite{Leung:2013}; only some additional terms arise in the latter because of the central binary's eccentric orbit. Both models work for $R \gtrsim 3\,a_{bin}$ and $e_{free}, e_{AB} \lesssim 0.1$. These conditions are partially fulfilled in TOI-1338 and TIC-172900988; the central eccentricities are larger than the above limits.

\subsection{Geometric approach}
\label{sec:geometric} % used for referring to this section from elsewhere

Another method for measuring the circumbinary orbital properties is defining geometric orbital elements \citep{Sutherland:2019}. These modified elements are calculated by averaging the maximum and minimum radial excursions of the circumbinary planet. In order to locate these distance extrema, a reasonable calculation window is required, and a specific value of the semi-major axis and eccentricity is found for each of these windows. The typical size of such windows is a bit larger than the Keplerian orbital period. We adopt the corrected definition by \cite{Bromley:2020}, which includes the forced motions by the binary system. A term $\Delta R_\pm$ is added accordingly, which accounts for the extrema of the radial excursions of the planet, in respect to the distance $R_P$. Assuming that the maximum and minimum radial excursions are nearly the same, an order of magnitude estimate of $\Delta R_\pm$ is (eq. (6) in \cite{Bromley:2020}):
 \begin{equation}\label{9}
\Delta R_\pm \sim R_P \frac{m_A m_B}{m_{bin}^2} \frac{a_{bin}^5}{R_P^5}\left(\frac{9}{4}+\frac{3n_S}{2n_{syn}}\right) \left(\frac{n^2_{bin}}{4n^2_{syn}-v_e^2}\right)\,.
\end{equation}
%
%In fact, $\Delta R_+$ and $\Delta R_-$ equal each other only if the binary masses are the same. In our case, the ratio of the central masses is ?? for TOI-1338 and ?? TIC-172900988, so we choose to adopt the above order of magnitude estimate (i.e. $\Delta R_+ \approx \Delta R_-$), rather than expanding  $\Delta R_+$ and $\Delta R_-$ (eq. (5) in \cite{Bromley:2020}).
%
Hence, the geometric mean distance is given by:
\begin{equation}\label{23}
a_{geo} = \frac{R_{max}+R_{min}-2\Delta R_\pm}{2}\,,
\end{equation}
and the geometric eccentricity is:
\begin{equation}\label{11}
e_{geo} = \frac{R_{max}-R_{min}-\Delta R_\pm}{2 a_{geo}}\,.
\end{equation}
$R_{max}$ and $R_{min}$ are the extreme values of the radial distance in a calculation window.

\subsection{Fourier analysis approach} \label{subsec:FFT}

The most robust approach to identify the frequencies of a motion due to oscillations is by applying a Fourier analysis. By evaluating the strength of a peak in a frequency spectrum, one can find the frequencies of a complicated motion. Inversely, this technique may be used to determine several orbital parameters. This has been done in \cite{Woo:2020}, where periodograms are used to find the frequencies $v_e$ and $v_i$ and the eccentricity $e_{free}$ of the circumbinary planets Kepler-16 b, Kepler-34 b, Kepler-35 b and the moons around Pluto and Charon. The resulting numbers suggest good agreement between the analytic and numerical methods for the exoplanets, and also compatible values with the orbital fitting of \cite{Showalter:2015} for Pluto-Charon.

\subsection{Keplerian elements approach} \label{subsec:Keplerian}

The usage of Keplerian elements is mentioned in \cite{Bromley:2020}. The following expressions give the semi-major axis and eccentricity, respectively, in orbits around a single mass:
\begin{equation}\label{24}
a_{Kep}=-\frac{Gm_{bin}m_P}{2E}\,,
\end{equation}
%
%\begin{equation}\label{13}
%e_{Kep}=\sqrt{1+\frac{2L^2E}{G^2m^2_{bin}m_P^3}}\,,
%\end{equation}
%
where $m_P$ is the planet's mass and $E$ is the mechanical energy. %and $L$ is the magnitude of the angular momentum. 
In circumbinary orbits though, these two quantities are not conserved, so eq.~(\ref{24}) may only be used only in the limit of orbits far away from the central binary. 

\subsection{Other methods} \label{subsec:other}

Some other ways of characterizing orbits around binary systems may be found in literature. For example, \cite{Georgakarakos:2015} derived analytical expressions for the eccentricity of a small circumbinary body. This solution is valid for low orbital eccentricities and includes a post-Newtonian relativistic correction. This model has already been validated for TOI-1338 and TIC-172900988 in \cite{Georgakarakos:2022}. The author found close values for the binary pericenter precession rates in the respective systems with \cite{Kostov:2020} and \cite{Kostov:2021}. Therefore, we do not explore this theory further here and we focus hereafter on the above methods.

\section{Results} \label{sec:sec4}

To explore the orbits in both systems, we employ a gravitational n-body code, which integrates the motions of the binary stars and the planets. The code, running in a Python/MATLAB environment, was adapted from an n-body simulation package created by Philip Mocz. We have used it in the past, in order to perform similar calculations concerning the circumbinary system of Pluto and Charon \citep{Gakis:2022}. 
Specifically, the leapfrog integration scheme is used to solve the differential equations of motion, involving the 'kick-drift-kick' technique to obtain positions and velocities. The velocity vector is updated at the kick half-steps of the integration, while the position is updated at the drift step. This ensures energy conservation, and time reversibility, as the leap-frog is a symplectic integrator. The gravitational force and the ensuing acceleration terms are computed before each kick-step, involving velocity updates.

It becomes obvious that results are sensitive to the choice of the timestep between the calculation moments. Therefore, caution is needed when choosing its value. The timestep should be small enough to minimize computational systematic errors, but sufficiently large to maintain calculation times below reasonable and practical limits as well. Here, unless otherwise stated, we assign the value of 8,500 s for TOI-1338 system and 11,500 s for TIC-172900988 ($\sim$150 times lower than the binary orbital period of the central orbital periods of both systems) as the fixed timestep of the n-body integrations. We have done numerical tests with smaller timesteps (200 times lower than the binary orbital periods), finding maximum deviations of $<$ 0.1\% in the final position of the planets.

The code treats all objects simulated as point-masses. We ratify this option as the characteristic size of the stars in the two systems lies in the range of $\sim$1-5\% of the binary separation, thus they can safely be treated as particles. Additionally, the gravitational interactions are modelled as purely Newtonian, neglecting the effects of general relativity or tidal evolution. The contribution of general relativity in the argument of periapsis precession rate of TOI-1338, 0.0005715$^\circ$ per cycle, is 0.0001132$^\circ$ per cycle and 0.0000055$^\circ$ per cycle is the contribution of tides \citep{Kostov:2020}. Besides, \cite{Kunovac:2020} showed that the central stellar orbits have not pseudo-synchronized, as their estimated lifetime is quite lower than their circularization timescale. The best-fitting model of \cite{Kostov:2021} yielded 1.70 $\times$ 10$^{-4}$ $^\circ$ per cycle for general relativity and 2.32 $\times$ 10$^{-5}$ $^\circ$ per cycle for tidal bulges, compared to the total change in the binary’s argument of periastron 2.88 $\times$ 10$^{-3}$ $^\circ$ per cycle, as for TIC-172900988. \cite{Georgakarakos:2022} found compatible results with these numbers. Evidently, the characteristic timescales of general relativity and tides are longer than the ones examined here, and we can safely ignore them for the purpose of our analysis. As an indication of the accuracy of the code, we keep track of the total mechanical energy of the whole dynamical simulated system, and find that it is well preserved.

As initial conditions of the n-body simulations, we assign the 3-dimensional vectors of positions and velocities in the lower parts of Tables \ref{tab:1} and Tables \ref{tab:2}, as given by \cite{Kostov:2020} and \cite{Kostov:2021}, respectively. These values of the Cartesian coordinates $x$, $y$, $z$ and projections of velocities $\upsilon_x$, $\upsilon_y$, $\upsilon_z$, came out of the best-orbital-fitting model.

\begin{figure}
    a\includegraphics[width=\columnwidth]{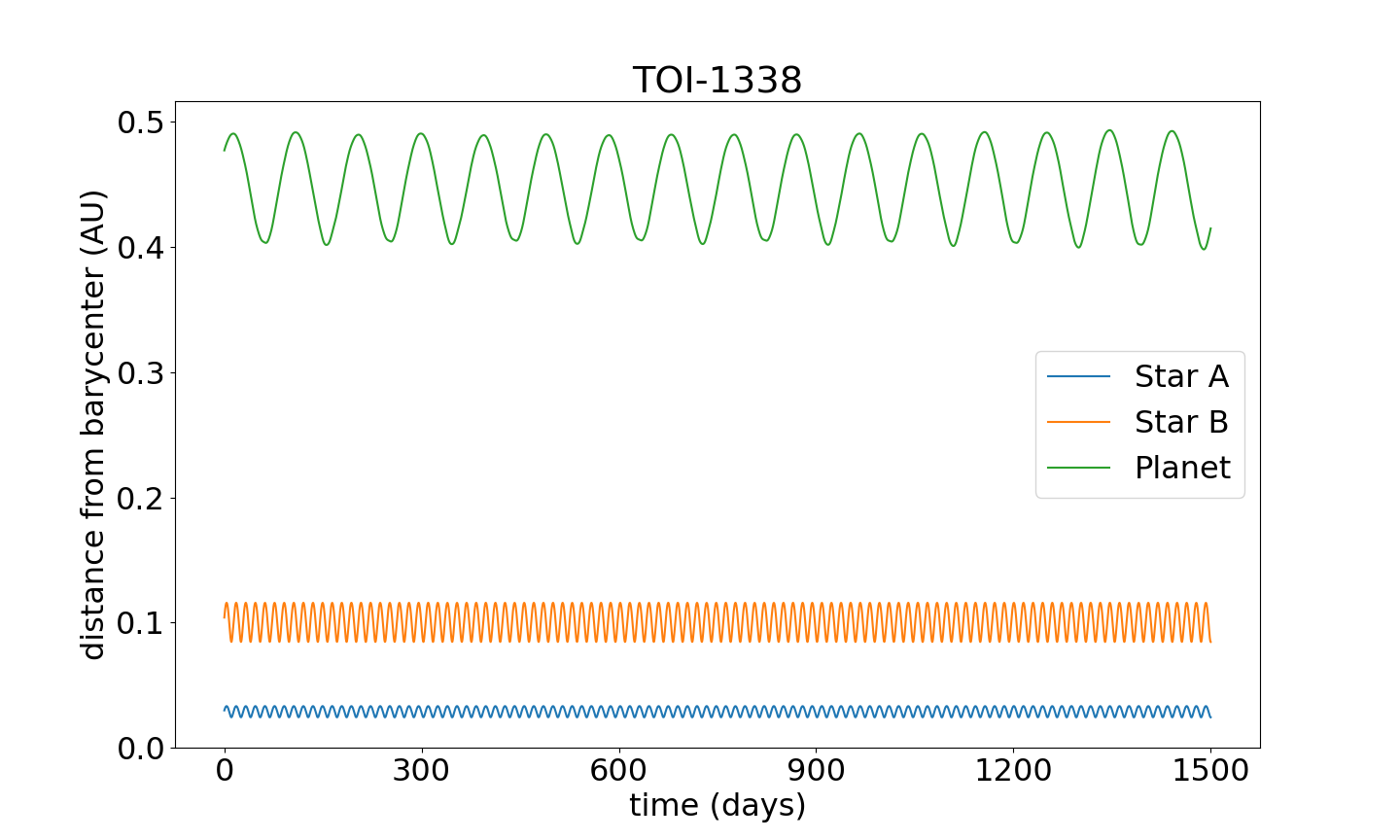}\\
    b\includegraphics[width=\columnwidth]{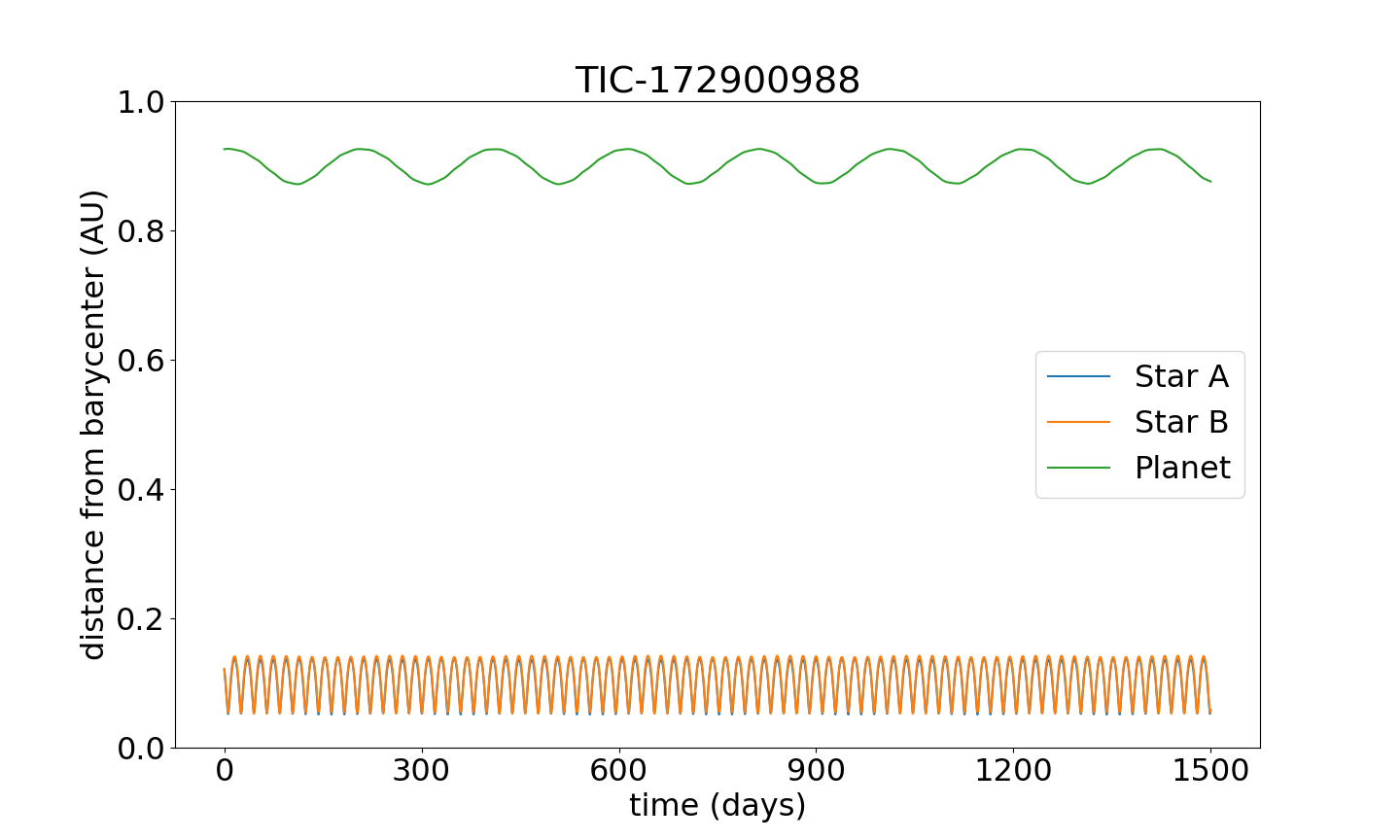}\\
\caption{Distance from barycenter through time for all three objects in each system. The systems were let to evolve for 1500 days. The three-dimensional vectors at the lower parts of Tables \ref{tab:1} and Tables \ref{tab:2} were used as initial conditions.}
\label{fig:1}
\end{figure}

As a first step, we integrate the 3-body gravitational systems of TOI-1338 and TIC-172900988 (separately) for 1500 days. The variation of distances for all objects constituting the systems through time for this timespan are given in Fig. \ref{fig:1}a and \ref{fig:1}b. We observe that the stars in TIC-172900988 spend most of their time near the apastron, as expected by their relatively high orbital eccentricity. Subsequently, we focus on the planetary orbits of both systems.

\begin{table*}
%\rotate
	\centering
%\tablecolumns{3}
\caption{Coefficients of the linearized model \label{tab:3}}
\begin{tabular}{clll}
%\tablewidth{0pt}
%\tablehead{  
%\colhead{Parameter\tablenotemark} & 
%\colhead{TOI-1338}   & 
%\colhead{TIC-172900988}   & 
%\colhead{} }
   Parameter & TOI-1338 & TIC-172900988 \\
%\startdata 
\hline
$C_0$ &$ 9.239440554847411E-05$ & $ 8.576602181685617E-05$ & \cr 
$C_1^0$ &$ -0.0001465707759020872$ & $ -9.543282714829757E-07$ & \cr 
$C^2_0$   &$ -0.00032917770419736925$ & $ -0.0001064386614728168$ & \cr 
$C^3_0$  &$ -2.3957145387709194E-05$ & $ -1.5598594608513152E-07$ & \cr 
$C_1^+$ &$ 2.5887414515180475E-06$ & $ 4.8388223594384444E-08$ & \cr 
$C_2^+$  &$ -1.8845317845581126E-05$ & $ -1.7493352891823355E-05$ & \cr 
$C_3^+$  &$ -3.0715292463349174E-06$ & $ -5.741239391102208E-08$ & \cr 
$C_1^-$   &$ 0.03100822432032993$ & $ 0.0017924989326466283$ & \cr 
$C_2^-$  &$ 0.0011489362113189293$ & $ 0.0010665114146381639$ & \cr 
$C_3^-$  &$ 5.471317543674639E-05$ & $ 1.0226874395044666E-06$ & \cr 
%\enddata
\hline
	\end{tabular}
\end{table*}

\begin{table*}
%\rotate
\centering
\caption{The first major oscillatory frequencies, in ($2\pi$ days)$^{-1}$, for the two planets. The rest of the them are calculated similarly. \label{tab:4}}
\begin{tabular}{cllr}
\hline
Parameter & TOI-1338 & TIC-172900988 \\
\hline
$v_e$ &$ 7.598734358964858E-07$ & $ 3.6104433661806714E-07$ & \cr
$1\cdot n_{syn}$ &$ 0.05789360358352305$ & $ 0.0458695537845109$ & \cr 
$2\cdot n_{syn}$   &$ 0.1157872071670461$ & $ 0.0917391075690218$ & \cr 
$3\cdot n_{syn}$  &$ 0.17368081075056915$ & $ 0.1376086613535327$ & \cr 
$1\cdot n_S - (1+1)\cdot n_{bin}$ &$ 0.1263573663222791$ & $ 0.09674697770205373$ & \cr 
$2\cdot n_S - (2+1)\cdot n_{bin}$  &$0.18425096990580217$ & $ 0.1426165314865646$ & \cr 
$3\cdot n_S - (3+1)\cdot n_{bin}$  &$ 0.2421445734893252$ & $ 0.1884860852710755$ & \cr 
$1\cdot n_S - (1-1)\cdot n_{bin}$   &$ 0.010570159155233011$ & $ 0.0050078701330319355$ & \cr 
$2\cdot n_S - (2-1)\cdot n_{bin}$  &$ 0.047323444428290035$ & $ 0.04086168365147896$ & \cr 
$3\cdot n_S - (3-1)\cdot n_{bin}$  &$ 0.10521704801181309$ & $ 0.08673123743598986$ & \cr 
\hline
	\end{tabular}
\end{table*}

\begin{table*}
%\rotate
	\centering
%\tablecolumns{3}
\caption{Free eccentricity estimates \label{tab:5}}
\begin{tabular}{clll}
%\tablewidth{0pt}
%\tablehead{  
%\colhead{Parameter\tablenotemark} & 
%\colhead{TOI-1338}   & 
%\colhead{TIC-172900988}   & 
%\colhead{} }
   Type of eccentricity & TOI-1338 & TIC-172900988 \\
%\startdata 
\hline
$e_{discov.}$ & $ 9.28291720948987292E-02$ & $ 2.73431915623053474E-02$ & \cr 
$e_{fitting}$ & $9.187908049138914E-02$ & $ 2.9170488497153143E-02$ & \cr 
$e_{geo}$ & $ 12.261706797008868E-02$ & $ 3.050522808337731E-02$ & \cr 
%\enddata
\hline
	\end{tabular}
\end{table*}

\begin{figure}
 a\includegraphics[width=\columnwidth]{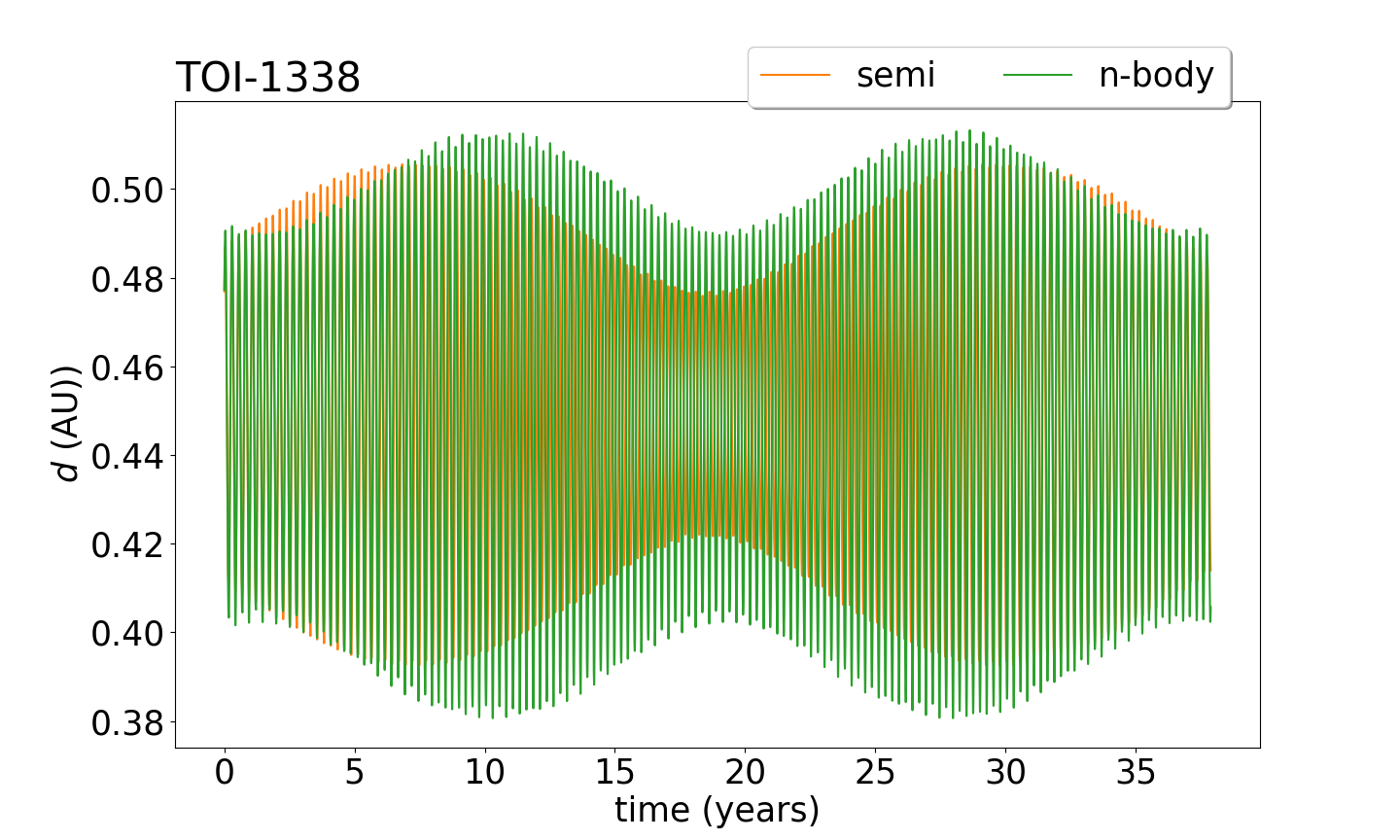}\\
    b\includegraphics[width=\columnwidth]{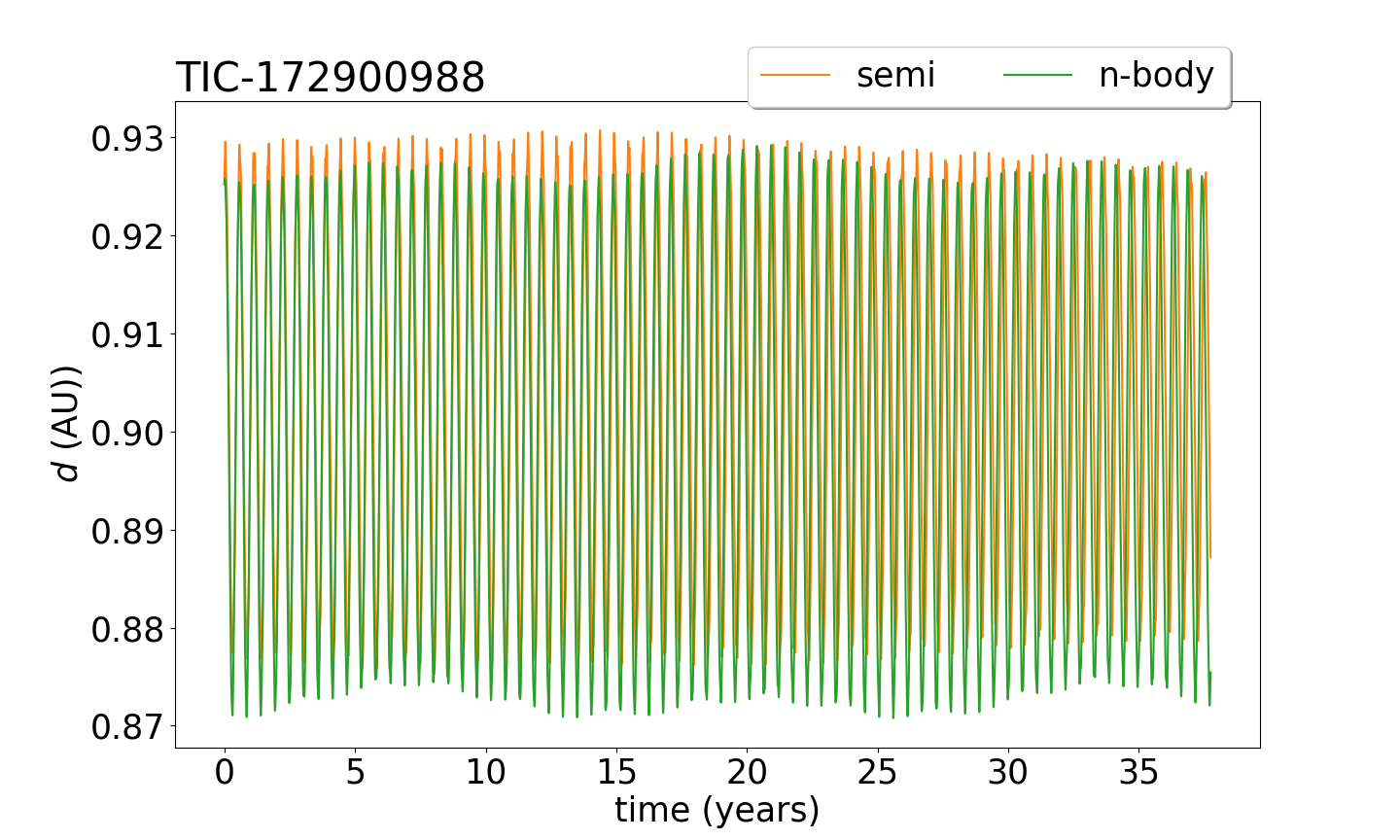}\\
\caption{Comparison of distances obtained using the n-body code (green plots) with semi-analytic distances obtained using eq.~(\ref{15}) (orange plots) through time for the planets objects in both system. The systems were let to evolve for about 40 years.}
\label{fig:8}
\end{figure}

Fig. \ref{fig:8}a and \ref{fig:8}b show the planetary distances from barycenter through about 40 years, as obtained by the n-body calculations (green plots) and the semi-analytic model (orange plots). The semi-analytic distance is calculated as:

\begin{equation}\label{15} r(t)=\sqrt{R^2(t)+z^2(t)}\end{equation}

As for the value of the parameter $e_{free}$ in eq.~(\ref{6}), we assign the nominal eccentricities of the planets (Tables \ref{tab:1} and \ref{tab:2}). %The coefficients $C_0$, $C_k^0$ and $C_k^{\pm}$ used for the calculations are presented in Table \ref{tab:3}.

We infer that the numerical and the semi-analytic approach yield close results. To quantify their relative difference, we define the norm: 
\begin{equation}
\|\delta R \|=\frac{1}{(t_f-t_i)}   \int_{t_i}^{t_f} \frac{| R_{n} - R_{m} |}{\langle R \rangle}  \,dt 
\end{equation}

We find $\|\delta R \|=$  6.87\% for TOI-1338 and $\|\delta R \|=$ 1.37\% for TIC-172900988. This is a result of the approximations carried out in the \cite{Leung:2013} model (like e.g. ignoring the non-linear terms of the gravitational forces) and the fact that we inevitably considered a finite number of terms in the sum of eq.~(\ref{6}) (up until $k=3$). We notice that the \cite{Leung:2013} model gives more accurate results when the orbitting object is farther from its host binary, as it happens with TIC-172900988. For reference, the relative discrepancy of the two methods was found 0.2925\% (maximum) for simulations of the Pluto-Charon moon system in a previous work \citep{Gakis:2022}.

As it can also be noticed from Fig. \ref{fig:1}a and \ref{fig:1}b, the distances approach a sinusoidal form, but small variations are observed. These small fluctuations are a result of the perturbations by the binaries. TOI-1338 planet orbits at a distance roughly five times larger than the semi-major axis of the binary stars. In TIC-172900988 the ratio is even larger, $\sim$9:1. At distances like these, the binary effect is not strong enough to induce powerful variations in the planetary orbits. Still, the binary systems do alter orbits beyond simple ones sufficiently described by Keplerian osculating elements. 

Because of this, the values of $R_P$ (loosely corresponding to the semi-major axis of the Keplerian orbits) and $e_{free}$ (loosely corresponding to the orbital eccentricity of Keplerian orbits) are not known with certainty. Thus, we fit them using non-linear least square differences between the n-body orbital distance patterns and eq.~(\ref{15}). Results yield $R_P=$ 0.4491255225049817 AU for TOI-1338 and $R_P=$ 0.89982619319084277 AU for TIC-172900988, confirming the accuracy of the semi-major axis in \cite{Kostov:2020} and \cite{Kostov:2021}, respectively. The fitting of $e_{free}$ results in 0.09187908049138914 as for TOI-1338 and 0.029170488497153143 for TIC-172900988. The additional terms in eq.~(\ref{15}) explain the deviation from the orbital eccentricities determined in \cite{Kostov:2020} and \cite{Kostov:2021}.

\begin{figure}
a\includegraphics[width=\columnwidth]{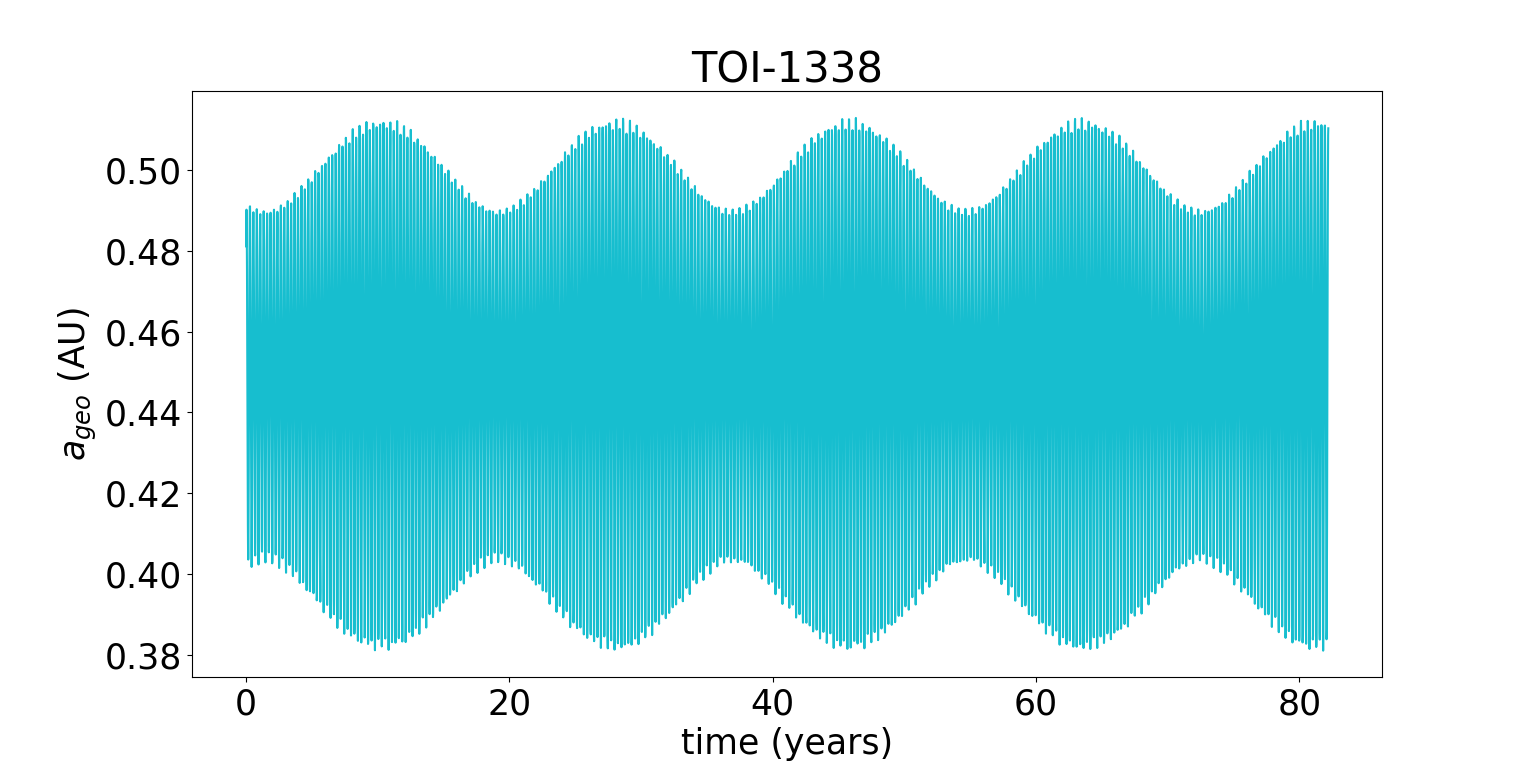}\\
    b\includegraphics[width=\columnwidth]{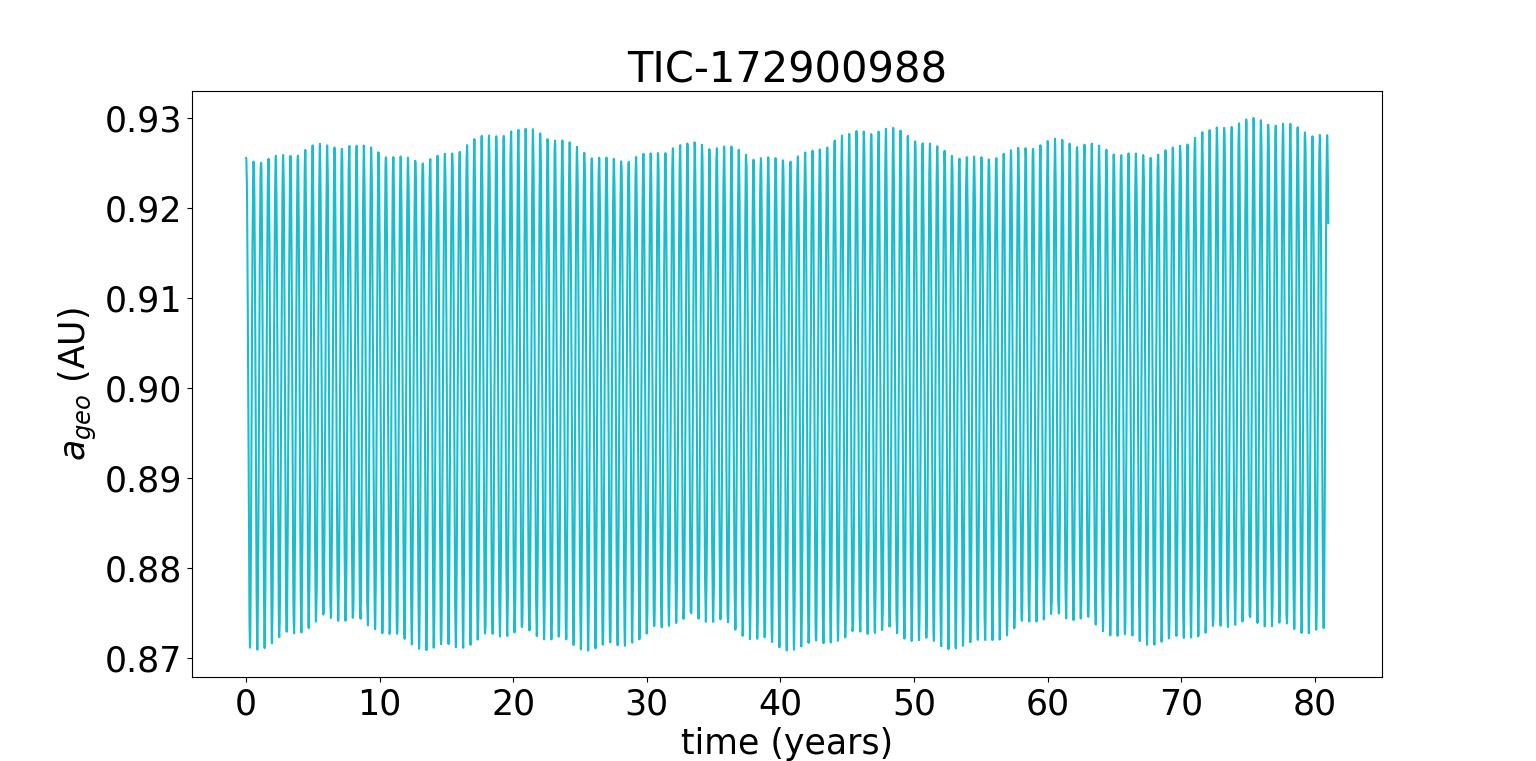}\\
\caption{Geometric mean distance $a_{geo}$ (eq.~(\ref{23})) through time for the planets objects in both system. The systems were let to evolve for about 80 years.}
\label{fig:3}
\end{figure}

\begin{figure}
a\includegraphics[width=\columnwidth]{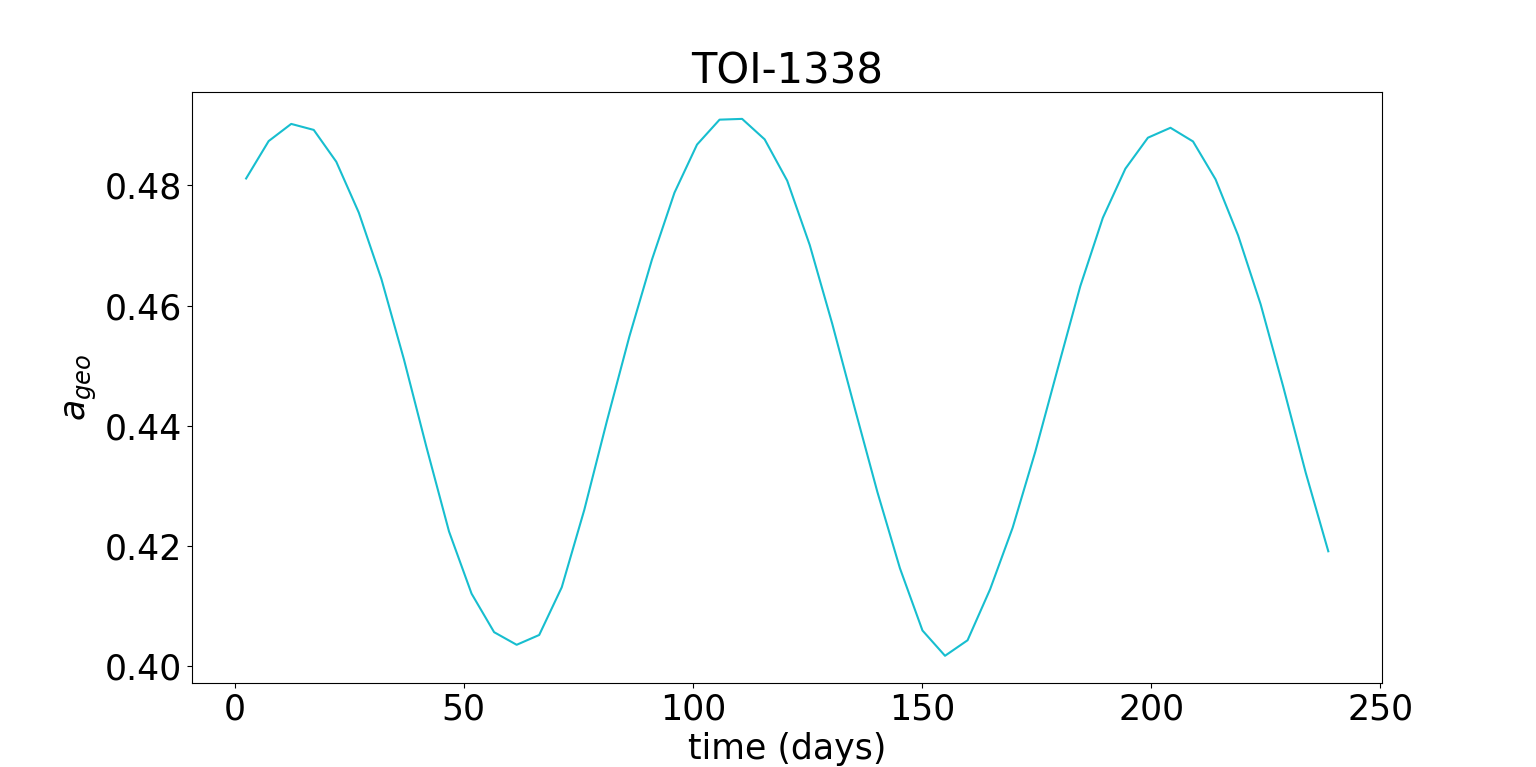}\\
    b\includegraphics[width=\columnwidth]{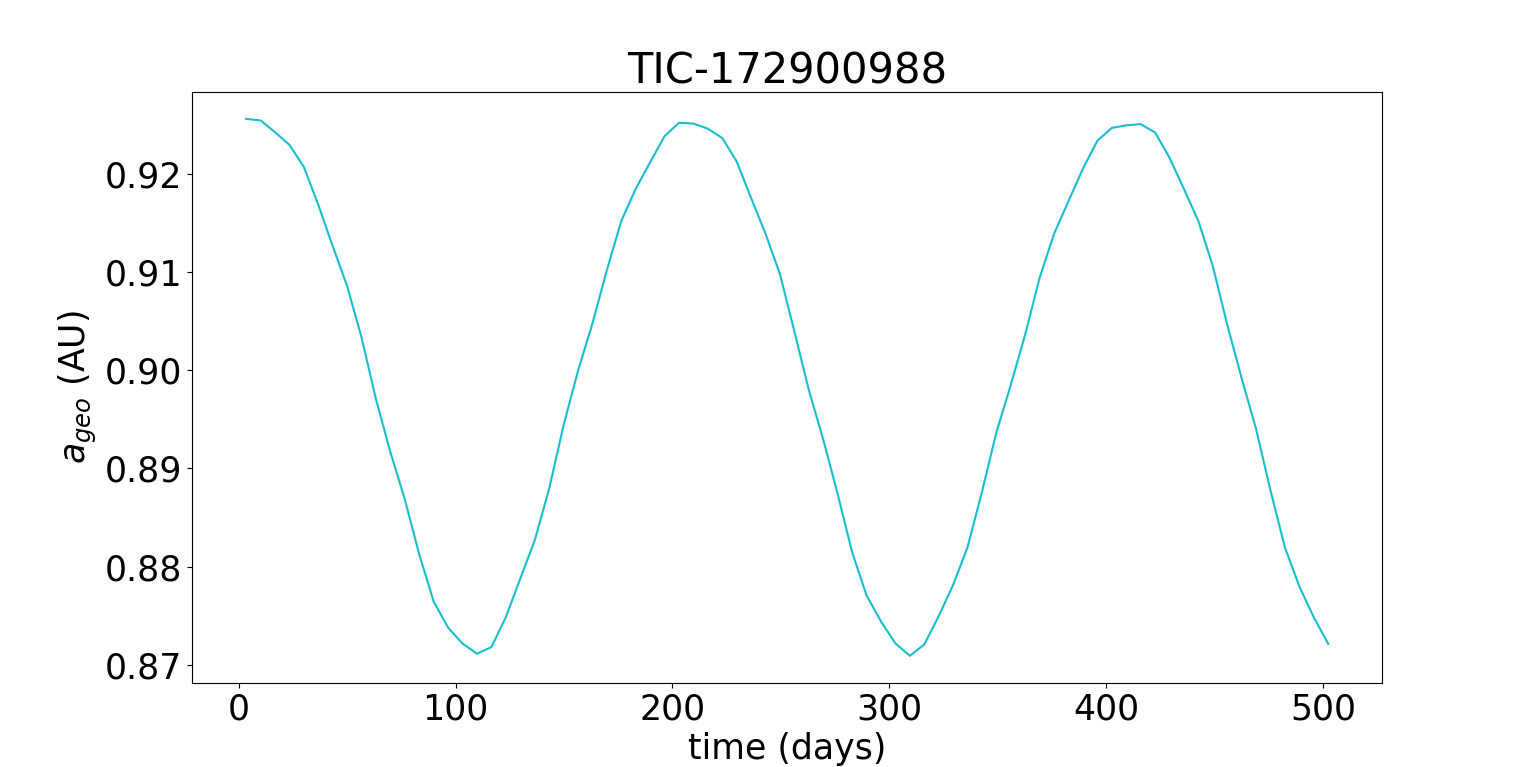}\\
\caption{Geometric mean distance $a_{geo}$ (eq.~(\ref{23})) through time for the planets objects in both system through a short period of time ($\sim$2.5$P$).}
\label{fig:20}
\end{figure}

We continue with implementing eq.~(\ref{23}) to get the evolution of the geometric distance $a_{geo}$ through time. This is shown in Fig. \ref{fig:3}a and \ref{fig:3}b through a a timespan of 80 years, in order to visualize the long-term variations $a_{geo}$. First, the systems were let to evolve for about 80 years. Then, we iterated the maximum and minimum planetary distance from the barycenter for calculation windows of $\sim$5 days and calculated $a_{geo}$ with eq.~(\ref{23}). Fig. \ref{fig:3}a and \ref{fig:3}b clearly demonstrate a major oscillation of larger period than the orbital periods. The images look like a superposition of many oscillatory modes, which we further discuss later. We also provide the variation of $a_{geo}$ through a shorter period of time (Fig. \ref{fig:20}a and \ref{fig:20}b), a rolling window of $\sim$2.5$P$, where $P$ is the Keplerian period of the planets (Tables \ref{tab:1} and Tables \ref{tab:2}), as suggested by \cite{Sutherland:2019}. Obviously, a quite low calculation window would give us almost identical plots with the n-body results. However, we advocate this method even if many samples of the positions through time are not available. In other words, the geometric method is a feasible tool for studying circumbinary orbits, and its accuracy increases with the number of available orbital data through time.

\begin{figure}
a\includegraphics[width=\columnwidth]{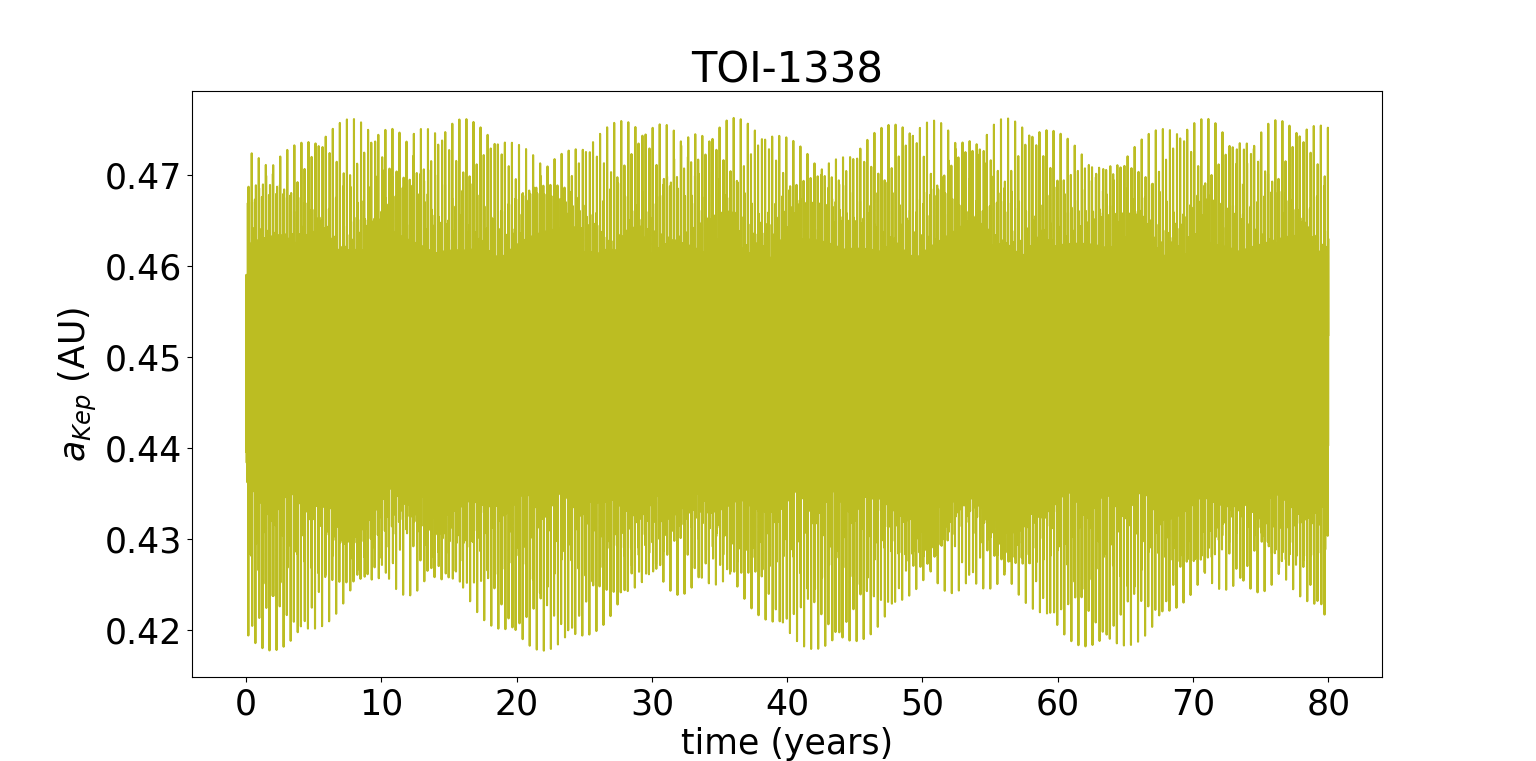}\\
    b\includegraphics[width=\columnwidth]{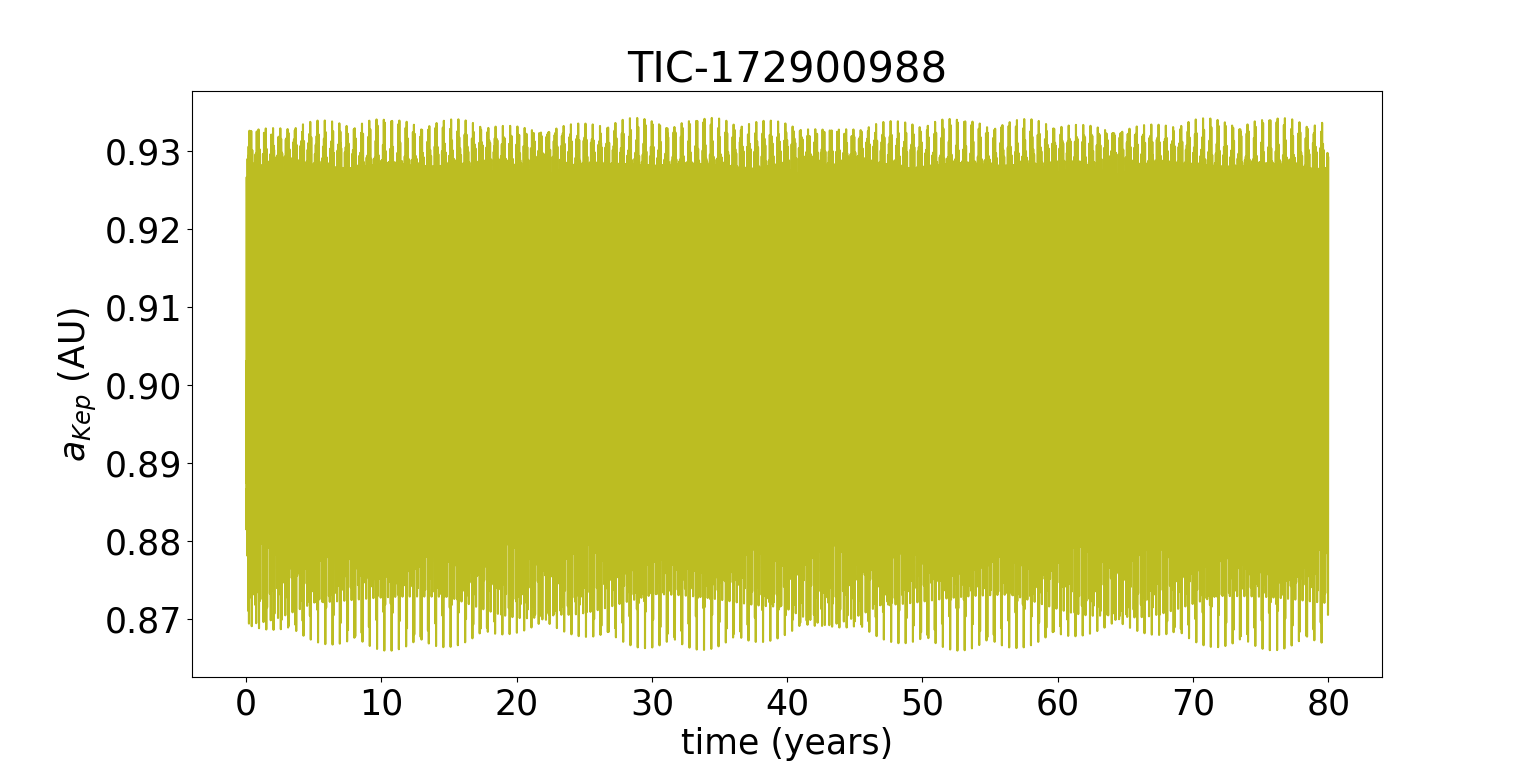}\\
\caption{Keplerian distances $a_{Kep}$ (eq.~(\ref{24})) through time for the planets objects in both system. The systems were let to evolve for 80 years.}
\label{fig:2}
\end{figure}
This is not true for the Keplerian method (Fig. \ref{fig:2}a and \ref{fig:2}b). As we mentioned earlier, Keplerian elements are not suitable for describing circumbinary orbits. Consequently, eq.~(\ref{24}) cannot be usually used to keep track of the distance of a circumbinary body. This is evident in Fig. \ref{fig:2}a and \ref{fig:2}b. Collating the values in the y axis (orbital distance) of the plots with the geometric and the Keplerian method suggests that values are distinct for the case of TOI-1338. Keplerian distances in TOI-1338 system are lower than the geometric distances, which suggests the non-conservation of the orbital energy. The plots are somewhat closer for TIC-172900988. This is explained by the fact that the planet in TIC-172900988 is more separated from its central binary than TOI-1338, thus the effect of the non-axisymmetric central potential is weaker.

For both exoplanetary systems though, the orbital patterns of the geometric and the Keplerian methods are not the same. $a_{Kep}$ demonstrates many fluctuations, more frequent and more irregular than the ones that $a_{geo}$ presents. Again, TIC-172900988 shows a better convergence than TOI-1338, justified by the proximity of TOI-1338 planet with its host stars.

Free eccentricity is just as important as orbital distance to understanding the nature of circumbinary orbits. For a more compelling comparison between the orbits deduced by observations ("ground truth" orbits) and the theoretical models, we include the eccentricity as an additional useful estimator. Specifically, we calculate a mean value of $e_{geo}$, implementing eq.~(\ref{11}) over the integration timespan of 80 years. The resulting values for the two planets are presented in Table \ref{tab:5}, along with the eccentricity determined in the "ground truth" orbits ($e_{discov.}$) and the fitted values using least square differences between the n-body integrations and the \cite{Leung:2013} model ($e_{fitting}$). Regarding TOI-1338, this eccentricity estimator shows a significant difference ($e_{geo}$ is about 32\% larger than $e_{discov.}$), which is attributed to the inconsistency of Keplerian orbits with circumbinary ones. TIC-72900988, being much farther relatively from the barycenter, appears to have closer values for the estimator ($\sim$12\% deviation). As noted in \cite{Bromley:2020}, the geometric eccentricity is a good approximation of the actual one, provided that both the binary and the planetary orbit eccentricities are much less than unity, while the binary radius is much less than the periastron distance, $(1-e)a_{p}/ a_{bin}\gg 1$, where $a_p$ and $a_{bin}$ are the planetary and binary semi-major axes, respectively, and $e$ is the planetary eccentricity. The eccentricities of TOI-1338 are 0.156 and 0.0928 for the binary and the planetary orbit respectively. In the system TIC-172900988 the binary and planetary eccentricities are 0.448 and 0.0273 respectively. The ratios of the periastron distance to the binary semi-major axis are 3.15 for TOI-1338, and 4.7 for TIC-172900988. Overall, the eccentricities are not sufficiently low, while the ratios of the periastron distance to the binary semi-major axis are not much larger than unity to allow for the linear theory to hold accurately. Therefore, deviations in the geometric eccentricity are indeed expected.

\begin{figure}
\includegraphics[width=\columnwidth]{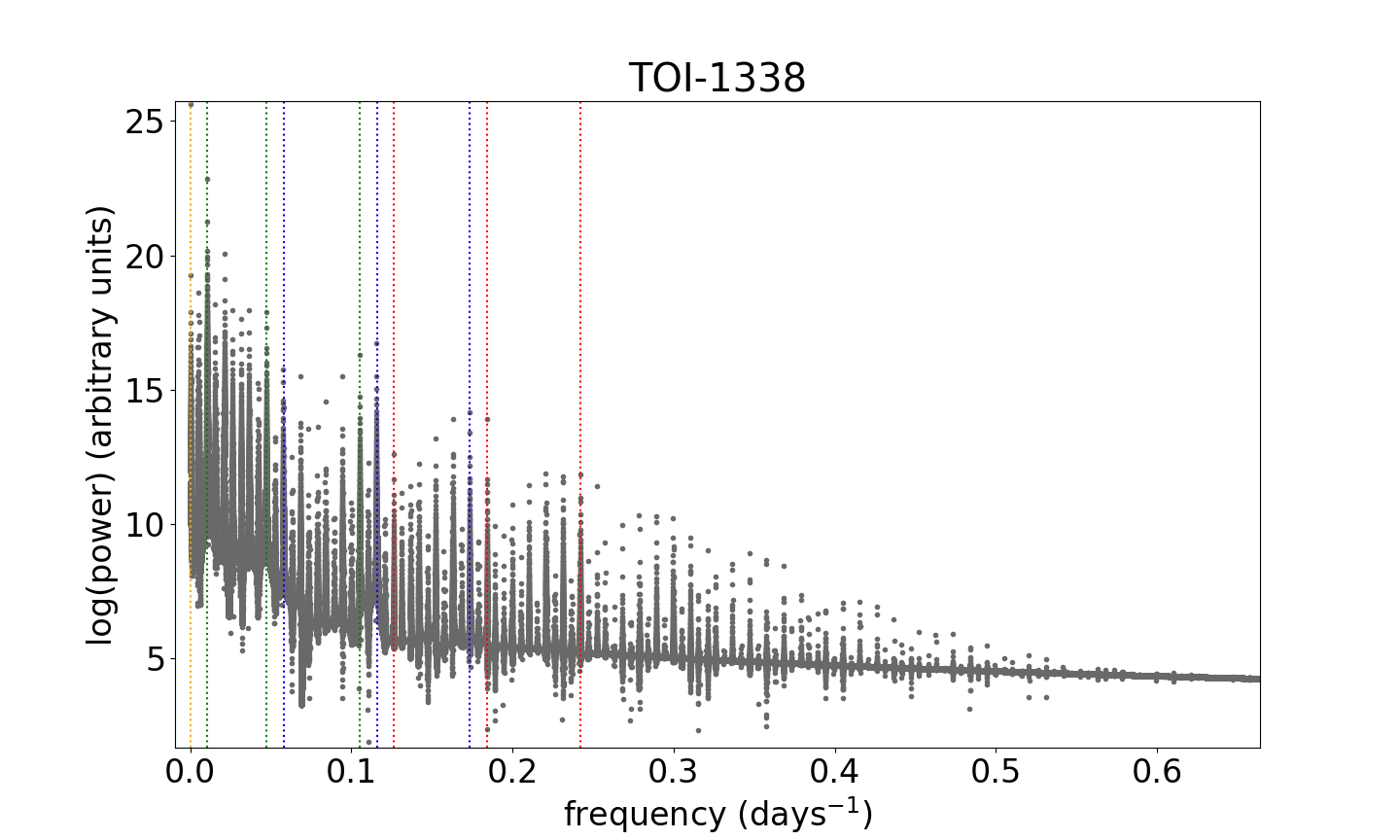}
\caption{The frequency power spectrum of the planet in TOI-1338 system. Simulations were let to evolve for 5,000 years. The orange vertical line corresponds to the frequency of $v_e$, whereas the three blue, red and green vertical lines are the first three terms ($k=1,2,3$) of $k\cdot n_{syn}$, $|k\cdot n_S - (k+1)\cdot n_{bin}|$ and $|k\cdot n_S - (k-1)\cdot n_{bin}|$ , respectively (as calculated in Table 2).}
\label{fig:6}
\end{figure}

\begin{figure}
\includegraphics[width=\columnwidth]{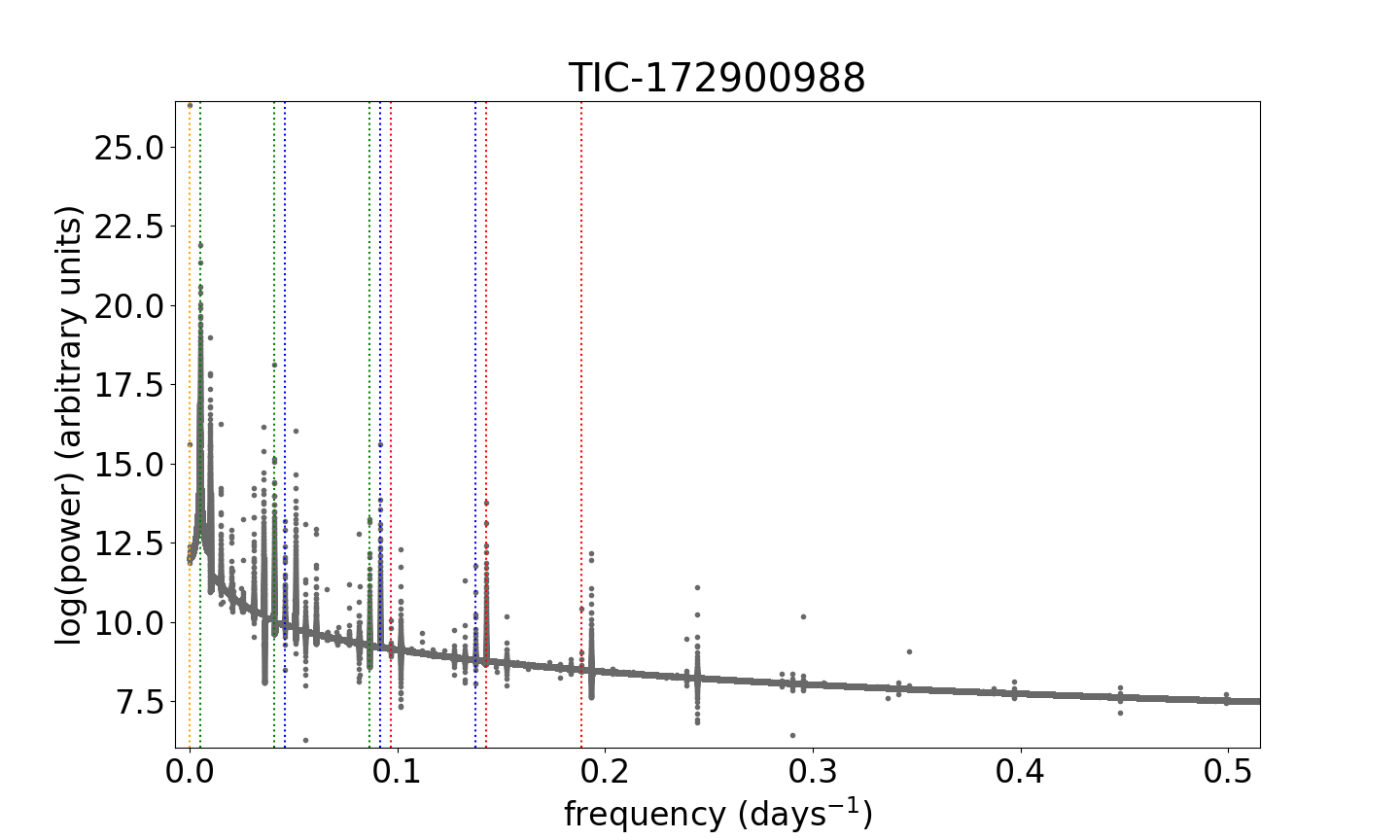}
\caption{The frequency power spectrum of the planet in TIC-172900988 system. Simulations were let to evolve for 5,000 years. The orange vertical line corresponds to the frequency of $v_e$, whereas the three blue, red and green vertical lines are the first three terms ($k=1,2,3$) of $k\cdot n_{syn}$, $|k\cdot n_S - (k+1)\cdot n_{bin}|$ and $|k\cdot n_S - (k-1)\cdot n_{bin}|$ , respectively (as calculated in Table 2).}
\label{fig:7}
\end{figure}

In order to decompose the oscillations, which give the orbital patterns when combined, we proceed to Fourier transformations. That way, we manage to infer the main oscillatory frequencies and compare the power of each one. The frequency spectra (Fig. \ref{fig:6} and \ref{fig:7}) were created by applying Fast Fourier Transformations (using the \texttt{scipy} library \texttt{fft}) in the orbital distances, after they were simulated for some thousands of years. Each vertical line represents an oscillatory mode. In fact, a sheer number of frequencies appear for both systems, which implies how complicated circumbinary orbits are. At a relatively farther orbital distance, TIC-172900988 experiences fewer oscillations than TOI-1338, which is greatly affected by the central binary orbits.

The colored vertical lines in Fig. \ref{fig:6} and \ref{fig:7} stand for the major frequencies anticipated by the linearized model of \cite{Leung:2013}, as it was outlined in Section \ref{subsec:linearized}. The orange vertical line corresponds to the frequency of $v_e$, whereas the three blue, red and green vertical lines are the first three terms ($k=1,2,3$) of $k\cdot n_{syn}$, $|k\cdot n_S - (k+1)\cdot n_{bin}|$ and $|k\cdot n_S - (k-1)\cdot n_{bin}|$, respectively (as calculated in Table \ref{tab:4}). The remaining peaks are expected by the theoretical model too. Specifically, if ones expands the terms in eq.~(\ref{15}), frequencies like $v_e$, $n_{bin}$, $n_S-n_{bin}$, $k\cdot n_S - (k+1)\cdot n_{bin}$, $k\cdot n_S - (k-1)\cdot n_{bin}$, $v_i$, combinations of them (e.g. sums, differences) and their harmonics appear. In other words, each vertical line corresponds to a frequency term of the semi-analytic model. Determining each one of them and matching it with its theoretical value is beyond the scope of this paper. Nevertheless, the fact that the basic lines noted in the spectra of Fig. \ref{fig:6} and \ref{fig:7} resemble some actual anticipated peaks validates our point. 

%\subsection{TIC-172900988} \label{subsec:resultsTIC}

We have to remark once again that our results were obtained adopting the currently accepted values for the orbital elements in the two exoplanetary systems. Of course, it is possible that future observations could provide us with orbital solution refinements. However, this possibility does not affect our analysis here, as our purpose was neither to provide new information about the orbital elements of the two systems nor to examine the observational properties of them. Rather than that, we give a comprehensive theoretical analysis of the orbital behaviors. Hence, our obtained knowledge about circumbinary orbits and ways to describe them are still valid even after a potential reconsideration in the orbital values of the systems, and could be even employed to determine their dynamical properties more accurately.

\section{Discussion} \label{sec:sec5}

%\subsection{Comparisons} \label{subsec:comparisons}

%comparisons

\subsection{Stability} \label{subsec:stability}

In the long-term numerical simulations (100,000 years) in \cite{Kostov:2020} and \cite{Kostov:2021}, the orbital distances appear as nearly straight lines as a function of time (for both systems). The variations of the eccentricity and inclination are also not large enough to implicate instabilities. It is then concluded that there is no indication for any radical change of this situation and the dynamical systems are stable. The results of our simulations do not reveal any signs for instabilities either. Our timescales are a lot shorter though, but even when increasing them we do not observe a different behavior. 

In general, the stability of 3-body hierarchical systems is a problem well-studied in literature. Examples of relevant works include \cite{Dvorak:1986, Holman:1999, Pilat:2003, Quarles:2018}. An elementary rule for the stability of a system consisting a binary system separated by a distance $a_{bin}$ and of eccentricity $e_{bin}$, and an object in a P-type orbit of radius $R_P$ around their barycenter may be summarized as $R_P \gtrsim 3 a_{bin}$, meaning that there do not exist stable motions inner to a distance of a short multiple of the binary separation \citep{Schneider:1994}. This is also the limit that the linearized model of \cite{Leung:2013} gives reasonable results. 

A more precise stability limit was derived by \cite{Holman:1999}:
\begin{eqnarray}
\frac{a_{crit}}{a_{bin}} &=& 1.60 + 5.10e_{bin} -2.22e_{bin}^2\nonumber \\
&+&4.12 \mu - 4.27e_{bin} \mu - 5.09 \mu^2 + 4.61 e_{bin}^2 \mu^2\,,
\end{eqnarray}
where $a_{crit}$ is the critical semi-major axis and $\mu=m_B/M_{bin}$ is the mass ratio of the binary. This above limit assumes circular P-type orbits, but since the planetary eccentricities of TOI-1338 and TIC-172900988 are lower than 0.1, we can safely adopt it for our analysis. Later studies like \cite{Mardling:2001} include planetary eccentricities as well. We deduce that the planets in TOI-1338 and TIC-172900988 systems remain beyond the respective critical distances $a_{crit}$ during the entire time of the simulation, confirming the stability of them.

%Section 2.2 of Simonetti 2020, extension by Quarles and limits 
%note that stability and instability limits not definitely decisive, islands of instabilities, or islands of stability inside the limit (Quarles 2018)

\subsection{On the existence of an additional planet in TOI-1338} \label{subsec:extra}

\begin{figure}
\includegraphics[width=\columnwidth]{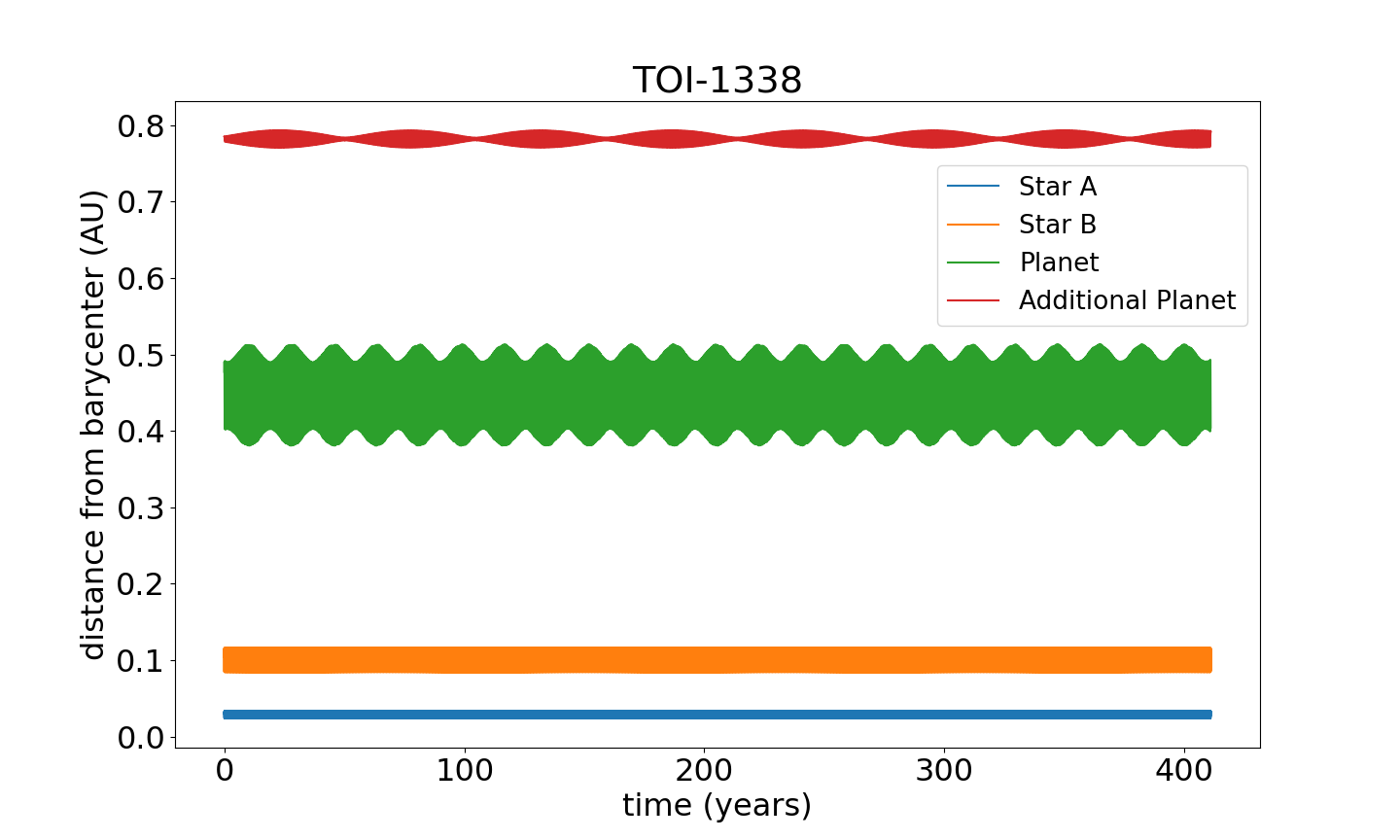}
\caption{Distance from barycenter through time for all four objects (the binary stars, the planet and the additional potential planet) in TOI-1338 system. The systems were let to evolve for 400 years.}
\label{fig:4}
\end{figure}

\begin{figure}
\includegraphics[width=\columnwidth]{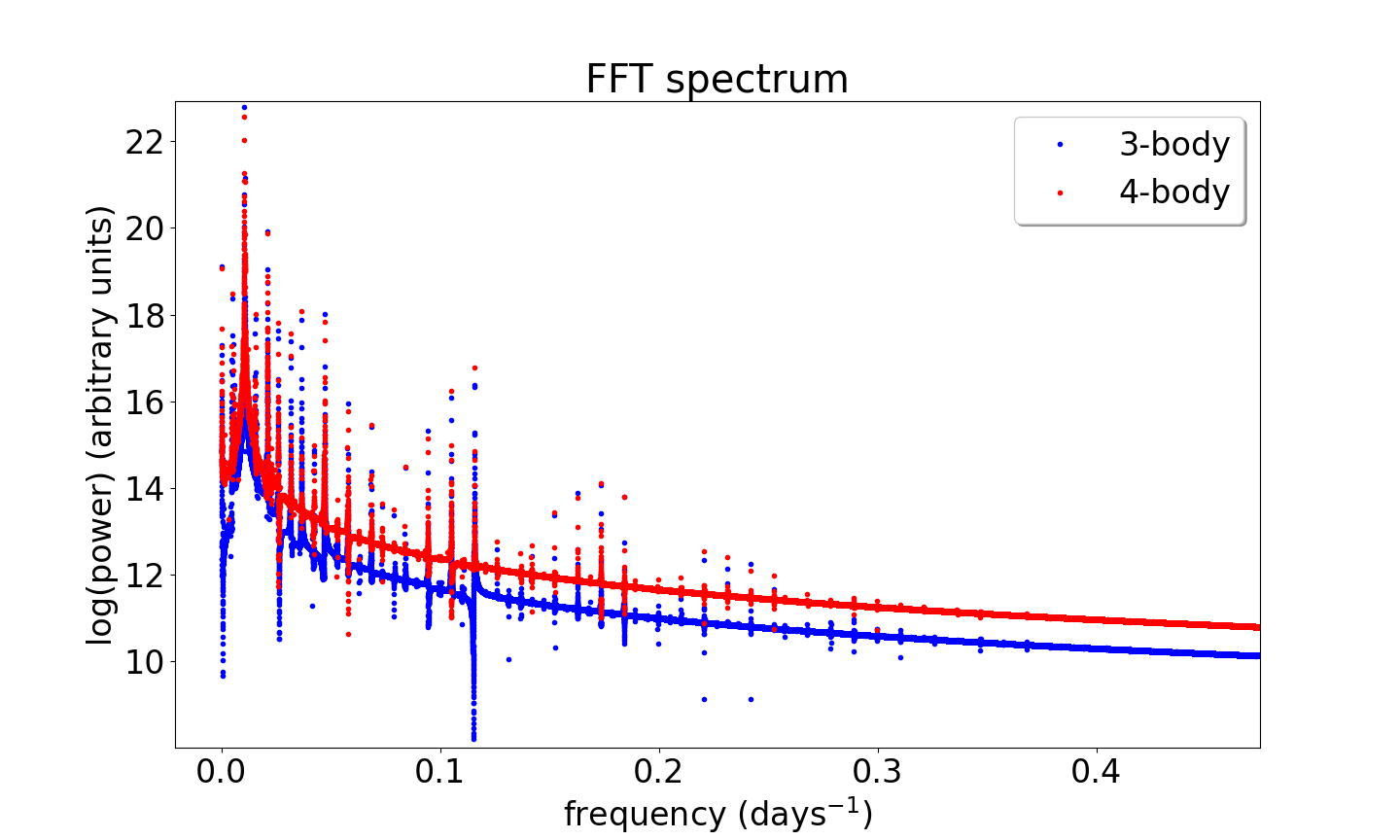}
\caption{The frequency power spectrum of the known planet in TOI-1338 system in the cases of the 3-body and 4-body systems.}
\label{fig:5}
\end{figure}

%Currently, only Kepler-47 is known to host multiple circumbinary exoplanets, three Neptune-sized ones \citep{Orosz:2012a}. 
Currently, only a single planet has been confirmed to orbit around the binary system of TOI-1338. A recent study, \cite{Standing:2021}, suggested the existence of a second planet in the system. Trying to place constraints in the mass uncertainty of the known planet, and after using radial velocity measurements by ESPRESSO and HARPS, a strong peak at a period of 220 days was detected. This second potential planet, at a computed mass of approximately 48 $M_{\earth}$, showed a trace of 0.1\% in the data periodogram, exceeding the random false alarm threshold. Additionally, the study of \cite{Standing:2021} set an upper limit of 12 $M_{\earth}$ for the confirmed planet’s mass. More data are needed to confirm the existence of this second planet.

Examining the hypothesis of a second candidate planet is not the objective of our study, but we test nonetheless the effects it could have in the dynamical system. We note that \cite{Georgakarakos:2022} added a fictional Earth-sized planet in their simulations of TOI-1338 and TIC-172900988, located in the habitable zone of the systems. As eccentricities did not rise throughout the numerical simulations, it was inferred that an extra Earth-like planet could not destabilize any of the systems.

We performed numerical simulations of the 4-body system in TOI-1338 (two circumbinary planets in motion around the barycenter of two host stars). The mass of the known was placed at 12 $M_{\earth}$ as suggested by \cite{Standing:2021}. The second (candidate) planet was given an initial velocity satisfying a zero orbital eccentricity. High-resolution spectroscopic measurements by \cite{Kunovac:2020} yielded a ratio of 600:4:1 for the angular momentum of the binary orbits, the planetary orbits and the primary star rotation in the TOI-1338 system. Furthermore, they are all aligned within a mutual inclination of 0.3$^\circ$. Hence, we initially placed the second planet in such an orbit, that its resulting angular momentum is parallel to the binary's. The distances from barycenter through 400 years for all four objects (the binary stars, the planet and the additional potential planet) in TOI-1338 system are given in Fig. \ref{fig:4}.

The plots in Fig. \ref{fig:4} do not recommend any radical changes in the orbital behavior of the bodies in TOI-1338 when adding the second planet. The larger period variations evident in the orbital patterns of the planet are a result of the central binary. For a more thorough inspection of the effects by the second planet, we provide the comparison of the frequency spectrum of the known planet in the 3-body and the 4-body situation (Fig. \ref{fig:5}). The major frequency peaks remain in the same positions in both cases. There are not any clearly visible formed modes induced in the 4-body case. Particularly, we can safely conclude that the extra planet does not cause any obvious instabilities.

More precisely, the mutual Hill radius  \citep{Chambers:1996} for $j$ and $q$ planets is given by:
\begin{equation} R_{H,jq}=\left(\frac{m_{j}+m_{q}}{3(m_{A}+m_{B})}\right)^{1/3}\overline{a} \,,
\end{equation}
where $\overline{a}$ is found by taking the average of the semi-major axes $a_j$, $a_q$;  $\overline{a}=(a_j+a_q)/2$. Let us assume, without any loss of generality, that $a_j > a_q$. To ensure stability, $a_j-a_q > \beta R_{H,jq}$ is required. The factor $\beta$ can range at minimum between 5 and 7 to ensure stability. This inequality was also used by \cite{Simonetti:2020} in order to examine the possibility of the existence of an additional planet within the boundaries of the habitable zone in circumstellar systems. The two planets in TOI-1338 are separated by $\beta = 13.7$, hence beyond the critical mutual distance for any instability. 

We note again that this conclusion does not imply that the second planet is evidently located in TOI-1338; we only investigate the dynamical implications this possibility would have and make comparisons with the single planet situation. More precise measurements are required to conclusively determine whether the system hosts two planets or not.

%Chen 2020 for planet-planet interactions

\section{Conclusions} \label{sec:sec6}

In this paper, we studied the orbits in the systems of TOI-1338 and TIC-172900988. Specifically, we analyzed the planetary circumbinary orbits and quantified the effect of the non-axisymmetric central potential, caused by the two host stars in the systems. We examined the short-term evolution of the motions for up to $\sim 10^4$ planetary orbits, numerically and theoretically in a more detailed way than previously, gave quantitative arguments about the suitability of each approach and discussed their implications. 

Circumbinary orbits cannot be approximated uniquely. For that reason, we compared several methods for such a task and provided a basis for future similar studies. At first, we applied the \cite{Leung:2013} model in practice for the first time for exoplanets orbiting around an eccentric binary star and determined the accuracy of such a model in the two dynamical systems. A reasonable approximation of the orbits was achieved by this semi-analytic model for orbits around eccentric binaries. The agreement is high for the first few tens of orbits but then the two methods get out of phase and the deviation scales with the planetary eccentricity. 

We demonstrated that geometric methods also give a good convergence, depending on the data available. However, adopting Keplerian elements to describe the circumbinary orbits is not realistic. In order to have an even more substantial dynamical analysis, we obtained an FFT analysis, decomposed the complicated circumbinary orbits into their components and identified the major frequencies. We verify that these frequencies are combinations of the oscillatory frequencies, as calculated using the \cite{Leung:2013} theoretical model.

We have additionally studied the stability of the systems, finding no signs of instabilities, so we now have strong evidence that both systems are stable. Finally, we have tested the impact an additional planet would have to the system. We have placed a 48 $M_{\earth}$ planet at a distance 0.8 AU in TOI-1338 system, as proposed by observations, and conclude that it would not significantly affect the dynamics of the system itself. 

Our study is not intended to restrict the uncertainties underlying the orbital elements of the two exoplanets. This is something that needs to be addressed in the future with further observations. Our purpose instead was to provide some necessary tools for a thorough theoretical analysis of circumbinary orbits and test their practicability and accuracy in practice for the exoplanetary systems of TOI-1338 and TIC-172900988, using the up-to-date available data. Therefore, we suggest that our conclusions could be employed for any system with similar properties.

\section*{Acknowledgements}

The numerical code used in this work was branched from an n-body code (https://github.com/pmocz/nbody-python) created by Philip Mocz. We thank an anonymous referee for their insightful comments.

%%%%%%%%%%%%%%%%%%%%%%%%%%%%%%%%%%%%%%%%%%%%%%%%%%
\section*{Data Availability}

The codes and the data that were used to prepare our models within the paper are available from the corresponding authors upon reasonable request.



\bibliographystyle{mnras}
\bibliography{example} % if your bibtex file is called example.bib

%\begin{thebibliography}{99}
%\end{thebibliography}

%%%%%%%%%%%%%%%%%%%%%%%%%%%%%%%%%%%%%%%%%%%%%%%%%%

%%%%%%%%%%%%%%%%% APPENDICES %%%%%%%%%%%%%%%%%%%%%

\appendix

\section{Tables}

\begin{table*}
%\rotate
\caption{Dynamical Parameters of the TOI-1338 system \label{tab:1}}
\begin{tabular}{rrrr}
\hline
  Parameter & Binary orbit & Planet orbit \\
\hline
$P$  (days)&$ 1.46085607280931704E+01$ & $ 9.51407742682822573E+01$ & \cr 
$a$  (AU) &$ 1.28783121829547487E-01$ & $ 4.49132971740733966E-01$ & \cr 
$e$     &$ 1.56005374830665650E-01$ & $ 9.28291720948987292E-02$ & \cr 
$i$  (deg) &$ 8.96576179191039415E+01$ & $ 8.92186126155212662E+01$ & \cr 
$\Omega$  (deg) &$ 0.00000000000000000E+00$ & $ 8.74517674648163101E-01$ & \cr 
$\omega$  (deg) &$ 1.17561331987513597E+02$ & $ 2.63336097695842795E+02$ & \cr 
true anomaly  (deg) &$ 1.11217395295404202E+02$ & $ 1.36038415640448932E+02$ & \cr 
mean anomaly  (deg) &$ 9.38816596366525289E+01$ & $ 1.28272133285056185E+02$ & \cr 
mean longitude  (deg) &$ 2.28778727282917799E+02$ & $ 4.00249031010939916E+02$ & \cr 
\noalign{\vskip 2mm}\hline\noalign{\vskip 2mm}
%\multicolumn{1}{c}{Parameter\tablenotemark} & \multicolumn{1}{c}{Star A} & 
%\multicolumn{1}{c}{Star B} & \multicolumn{1}{c}{Planet} \cr
Parameter & Star A & Star B & Planet \cr
\noalign{\vskip 2mm}\hline\noalign{\vskip 2mm}
Mass ($M_{\odot}$) & $ 1.03784970719363567E+00$ & $ 2.97388770751337850E-01$ & $ 9.06017229632760055E-05$ \cr 
$x$ (AU) & $ 1.95196590778876217E-02$ & $-6.82335050435495666E-02$ & $ 3.68709659800795730E-01$ \cr 
$y$ (AU) & $ 1.32648896371446897E-04$ & $-4.65900424342245798E-04$ & $ 9.75628498103004414E-03$ \cr 
$z$ (AU) & $ 2.22880532978689296E-02$ & $-7.78747286204819755E-02$ & $ 3.02645766911772141E-01$ \cr 
$v_x$ (AU day$^{-1}$) & $-7.66586168788931707E-03$ & $ 2.67578221143398472E-02$ & $-1.61533175726252566E-02$ \cr 
$v_y$ (AU day$^{-1}$) & $ 5.45640776560390306E-05$ & $-1.90441050119945319E-04$ & $ 6.31089229491667655E-05$ \cr 
$v_z$ (AU day$^{-1}$) & $ 9.13006908476301365E-03$ & $-3.18697187907663951E-02$ & $ 2.27034214206392367E-02$ \cr 
\hline
	\end{tabular}
\end{table*}

\begin{table*}
%\rotate
\caption{Dynamical Parameters of the TIC-172900988 system \label{tab:2}}
\begin{tabular}{rrrr}
\hline
  Parameter & Binary orbit & Planet orbit \\
\hline
$P$  (days)&$ 1.96581697466411960E+01$ & $ 2.00453455776095382E+02$ & \cr  
$a$  (AU) &$ 1.91925787812358140E-01$ & $ 9.02849883123379215E-01$ & \cr  
$e$     &$ 4.47857817550560366E-01$ & $ 2.73431915623053474E-02$ & \cr 
$i$  (deg) &$ 9.05720931926375528E+01$ & $ 9.18314640412624783E+01$ & \cr 
$\Omega$  (deg) &$ 0.00000000000000000E+00$ & $ 3.52795698161293025E-01$ & \cr 
$\omega$  (deg) &$ 6.95986966505942490E+01$ & $ 1.57842415515142477E+02$ & \cr  
true anomaly  (deg) &$ 2.14795791997321089E+02$ & $ 1.52599413335829297E+02$ & \cr 
mean anomaly  (deg) &$ 2.54508036598815352E+02$ & $ 1.48094540591452386E+02$ & \cr 
mean longitude  (deg) &$ 2.84384567423867907E+02$ & $ 3.45804791152572307E+02$ & \cr  
\noalign{\vskip 2mm}\hline\noalign{\vskip 2mm}
%\multicolumn{1}{c}{Parameter\tablenotemark} & \multicolumn{1}{c}{Star A} & 
%\multicolumn{1}{c}{Star B} & \multicolumn{1}{c}{Planet} \cr
Parameter & Star A & Star B & Planet \cr
\noalign{\vskip 2mm}\hline\noalign{\vskip 2mm}
Mass ($M_{\odot}$) & $ 1.23872067209601799E+00$ & $ 1.20195846007516960E+00$ & $ 2.84367305505320387E-03$ \cr 
$x$ (AU) & $-3.05117762048147702E-02$ & $ 2.97755496161732185E-02$ & $ 7.05634622502878561E-01$ \cr 
$y$ (AU) & $-1.18334732414564496E-03$ & $ 1.16402958706238267E-03$ & $ 2.34631694465425554E-02$ \cr 
$z$ (AU) & $ 1.16468463181218757E-01$ & $-1.18616173378369855E-01$ & $-5.97881635138625467E-01$ \cr 
$v_x$ (AU day$^{-1}$) & $-1.85668502216597038E-02$ & $ 1.90921472185390795E-02$ & $ 1.79954987672769490E-02$ \cr 
$v_y$ (AU day$^{-1}$) & $ 1.37117431317775707E-04$ & $-1.39996410566221895E-04$ & $-5.55734287033887329E-04$ \cr 
$v_z$ (AU day$^{-1}$) & $-1.36914505868482791E-02$ & $ 1.40608913970410084E-02$ & $ 2.08446965916781538E-02$ \cr  
\hline
	\end{tabular}
\end{table*}

%%%%%%%%%%%%%%%%%%%%%%%%%%%%%%%%%%%%%%%%%%%%%%%%%%

% Don't change these lines
\bsp	% typesetting comment
\label{lastpage}
\end{document}